\documentclass[%
superscriptaddress,
preprint,
 amsmath,amssymb,
 aps,
]{revtex4-2}

\usepackage{graphicx}
\usepackage{dcolumn}
\usepackage{array}
\usepackage{bm}

\usepackage{subcaption}
\usepackage{float}  
\usepackage{gensymb}
\usepackage{minted}
\usepackage{makecell}
\usepackage{amsmath}
\usepackage{amsthm}
\usepackage{orcidlink}
\usepackage{xcolor} 
\usepackage{tikz}
\usepackage{xurl}
\usetikzlibrary{arrows.meta, positioning}

\usepackage{hyperref}

\usepackage{natbib}
\usepackage{threeparttable}
\usepackage[section]{placeins}

\begin{document}


\title{Variability in a Model of Divergent and Convergent \\Coupled Ice Streams: Synchronisation, Transients and Chaos}

\author{Joshua Grimstead\,\orcidlink{0009-0007-0771-2157}}

\affiliation{Department of Mathematics and Statistics, University of Exeter, Exeter, U.K.}

\author{Peter Ashwin\,\orcidlink{0000-0001-7330-4951}}
\thanks{Contact author: Peter Ashwin}
\email{p.ashwin@exeter.ac.uk}
\affiliation{Department of Mathematics and Statistics, University of Exeter, Exeter, U.K.}
%
 \author{Tanja I. Schindler\,\orcidlink{0000-0002-9056-8884}}
 \email{t.schindler@exeter.ac.uk, tanja.schindler@uj.edu.pl}
\affiliation{Department of Mathematics and Statistics, University of Exeter, Exeter, U.K.}
\affiliation{Faculty of Mathematics and Computer Science, Jagiellonian University, ul.\ Łojasiewicza 6, 30-348 Krakow, Poland
}




\date{\today}

\begin{abstract}
Fast-flowing glaciers, known as ice streams, control the majority of ice sheet mass loss due to thermomechanical coupling at their base, which can induce basal lubrication and highly variable flow. Even quite simple models can show `binge-purge' oscillations. Recent work has highlighted that coupling between divergent ice streams can also give rise to chaotic variability. In this paper, we return to a model of ice streams with a divergent coupling to identify and correct some physical inconsistencies, most notably a global volume conservation error. We propose and validate a revised Divergent Coupling (DC) model that ensures rigorous mass balance. Bifurcation analysis and Poincaré sections reveal previously unidentified complex frequency-locking (Arnold tongues), a period-adding cascade and chaos associated with torus breakdown. 
We also consider an equivalent model with a Convergent Coupling (CC), where we find no evidence of chaotic variability.
These findings suggest the importance of specific topologies on ice stream variability. 
\end{abstract}
\maketitle

\tableofcontents

\section{Introduction} 

Ice streams (fast-flowing glaciers) are known to exhibit oscillatory behaviour via thermomechanical coupling to subglacial till, where meltwater lubrication triggers fast-flowing ``streaming" behaviour, followed by stagnation when the bed refreezes. Robel \textit{et al.} \citep{R13} present an idealised model (hereafter R13) of ice stream hydrology coupled to flow dynamics, capturing both variable ice thickness and meltwater content (the two degrees of freedom required for oscillatory behaviour), which has facilitated recent developments \citep{mantelli_stochastic_2016,mann_subtemperate_2025}.

A high resolution ice sheet model showed that a pair of retreating ice streams sharing a common upstream reservoir can suppress tipping of the Greenland Ice Sheet through chaotic transients \citep{K26b}. Inspired by this, Kypke \textit{et al.} \citep{K26} (hereafter K26) formulated a coupled model of Robel's ice streams with a similar divergent topology, discovering similar chaotic transients in this idealised configuration.

Since uncoupled ice streams exhibit binge-purge oscillations, Kypke \textit{et al.} \cite{K26} suggest it is possible that when coupled in a divergent topology, the system behaves as a pair of coupled limit cycle oscillators. This means that, for a relatively weak coupling, the attracting dynamics can be described as coupled phase oscillators on an invariant $2$-torus, while for a stronger coupling the $2$-torus can break up to give chaotic variability. 
As such, we would expect K26 to demonstrate phenomena such as phase locking and Arnold tongue behaviour similar to the Arnold circle map \citep{pikovsky_synchronization_2002}. Similar behaviour has been found where oscillatory processes interact \citep{crucifix2012oscillators,ashwin_chaotic_2018}, suggesting that phase locking on glaciological timescales is not unprecedented and likely a feature of such systems. 

The paper is structured as follows. In Section~\ref{sec:K26} we review the elements of the R13 and K26 models. In doing so, we highlight two errors in the K26 formulation. In particular, there is a lack of volume conservation in the coupling between the component `boxes' of the divergent ice stream. In Section~\ref{sec:correct}, we correct these to give a new DC model that ensures volume conservation and analyse the variability in the corrected model. Similar behaviour as in \cite{K26} is found, but, for example, there are different regions in the parameter space where chaotic behaviour occurs.
To better understand the structure of parameter space, in Section~\ref{sec:sync_chaos} we explore the dynamics of the DC model by varying two parameters - this involves mapping out regions of frequency synchronisation and chaotic behaviour, where indeed we find behaviour typical of breakdown of the $2$-torus for coupled phase oscillators to be responsible for the emergence of chaos. Section~\ref{sec:consequences} outlines some consequences for variability in coupled ice streams, including a comparison of the DC model with a similar model of ice streams with Convergent Coupling (CC), where we do not find chaotic behaviour. The paper finishes with some discussion in Section~\ref{sec:discuss} of the relevance of the results to ice streams and some challenges for future work.

\section{The Robel/Kypke ice stream model} 
\label{sec:K26}

The paper \cite{R13} presents a spatially lumped model, R13, that ignores explicit topology, describing only the temporal evolution of ice streams and modelling changes in ice thickness while assuming constant width and length. The model explores how ice stream dynamics depend on the subglacial heat budget, meltwater production, and till deformation through three till states (frozen, partially frozen, and thawed), with lubrication controlling basal motion. The model can be expressed explicitly per till state, as in K26 which is derived from the paper's associated MATLAB code \cite{robelcode} (hereafter R13c). Note that we use an adjusted notational convention to that of R13 for convenience and consistency; see Appendix \ref{appendix:param} for a record of the notation and specific parametrisations used in this work.

As till water content is the primary driver of the model's internal mechanics, we build the model framework up from this specific mechanism before describing ice stream behaviour for each till state. R13 uses the following equation to describe the change in water content $w$ when $w>0$, 
\begin{equation}\label{eq:water_content}
\frac{dw}{dt} = m - \frac{Q_d}{L W},
\end{equation}
where $m$ is melt rate, $L$ is ice stream length, $W$ is ice stream width and $Q_d$ is subglacial discharge. This is dependent on $w$, such that 
\begin{equation*}
Q_d =
\begin{cases}
0, & \text{if } w < w_s \text{ or } m < 0 \\
m L W, & \text{otherwise.}
\end{cases}
\end{equation*}
where $w_s$ is a water content saturation threshold. This implies that $\frac{dw}{dt}=0$ when the water content has surpassed the saturation threshold $w>w_s$, such that the total till volume is capped at $h_\mathrm{till,max}$. 
Basal melt rate $m$ is related to geothermal heat flux $q_g$, conduction into the ice $\frac{K_I (T_s - T_b)}{H}$, and frictional heating $\tau_f u_b$ through: 
\begin{equation*}
m = \frac{1}{\rho_I L_f}
\left[
\tau_f u_b + q_g - \frac{K_I (T_s - T_b)}{H}
\right],
\end{equation*} 
where $H$ is ice stream thickness, $\rho_I$ is the density of ice, $L_f$ is the latent heat of fusion, $K_I$ is the thermal conductivity of ice, $\tau_f$ is basal frictional stress, $u_b$ is basal velocity, $T_b$ is basal temperature and $T_s$ is surface temperature. Note that both temperature variables $T_b$ and $T_s$ are positive for negative temperatures. Here, `negative melting' $m <0$ corresponds to the water in the till freezing. This description of heat diffusion through the basal layer serves as an approximation to a fully dynamic model for heat diffusion \citep{macayeal_bingepurge_1993}. \\
\indent Basal frictional stress $\tau_f$ is modelled as a Coulomb friction law which assumes there is no sliding when the yield stress of the bed is not attained. This is expressed in R13c (and K26) in terms of the void ratio \citep{tulaczyk2000a},
\begin{equation*}
    \tau_f = \tau_0 \exp(-ce),
\end{equation*}
where $c$ is an empirical till coefficient and $\tau_0$ is an empirical till coefficient dependent on void consolidation threshold $e_c$ representing the maximum basal frictional stress \citep{tulaczyk2000a,robelcode}. The till void ratio is the ratio of water volume to solid volume in the till $e =\frac{V_w}{V_s}$ \citep{tulaczyk2000b}, which relates unfrozen till solid thickness and water content $w = e h_{\mathrm{till},s}$. 

Switching between the different till states is non-linear, and depends on two key variables, $w$ and $h_{\mathrm{till},s}$. In the frozen till state, $w=0$ and  $h_{\mathrm{till},s} = 0$, neither of which vary. When basal temperatures are sufficiently warm the till becomes only partially frozen, allowing for water in the till $w>0$ such that the unfrozen till solid thickness is both non-zero and less than its maximum, $0<h_{\mathrm{till},s}<h_{\mathrm{till,max}}$. When the till is fully thawed, we have $w>0$ and $h_{\mathrm{till},s} = h_{\mathrm{till,max}}$, allowing for basal sliding. We describe these switches in the non-linear dynamics in more detail below.

\subsection{Frozen till}

For a frozen till $w=0$, and consequently $\frac{dw}{dt}=0$ thus $\frac{de}{dt} =\frac{dh_{\mathrm{till},s}}{dt} = 0$ where $e = e_c$ and $h_{\mathrm{till},s} = 0$.
As there is no change in water content, all heat fluxes result in varying the basal temperature $T_b$, which is either at the melting threshold with a negative melting rate, which maintains the frozen state $T_b = T_m = 0$ and $m<0$, or the basal temperature is positive, $T_b>T_m = 0$. 
\begin{equation}\label{eq:Tb}
\frac{dT_b}{dt} = -\frac{\rho_I L_fm}{C_I \eta_b},
\end{equation}
where $C_I$ is the volumetric heat capacity of ice and $\eta_b$ is the thickness of the temperate ice layers. 
Note that in (\ref{eq:Tb}) the frictional heating term $\tau_f u_b = 0$ as the ice stream is stagnant with frictional stress $\tau_f = \infty$, such that $u_b=0$.
  
\subsection{Partially frozen till} 

For a partially frozen till, there is a small amount of water content $w>0$, thus equation (\ref{eq:water_content}) applies. However, the ice stream is stagnant $u_b = 0$ and $e = e_c$, thus $\frac{de}{dt} = 0$.
As such, the unfrozen till solid thickness is either less than its maximum $h_{\mathrm{till,max}}$ but not yet zero, thus $0 < h_{\mathrm{till},s} < h_{\mathrm{till,max}}$, or at its maximum but with negative melt rate $h_{\mathrm{till,max}}=h_{\mathrm{till},s}$, $m < 0$. Thus $\frac{dw}{dt} = e_c \frac{dh_{\mathrm{till},s}}{dt}$, such that 
\begin{equation}\label{eq:h_till}
   \frac{dh_{\mathrm{till},s}}{dt} = \frac{m}{e_c}.
\end{equation}
As all the energy from geothermal heating is preventing meltwater from changing state, the pressure melting temperature does not vary such that $T_b = T_m$ and $\frac{dT_b}{dt} = 0.$

\subsection{Thawed/unfrozen till} 

For a thawed/unfrozen till $w>0$, thus equation (\ref{eq:water_content}) applies and basal temperature is at the pressure melting point $T_b = T_m = 0$ and $\frac{dT_b}{dt}=0$. A thawed till also implies that the bed is saturated, thus $e > e_c$ and  $h_{\mathrm{till},s} = h_{\mathrm{till,max}}$ such that $\frac{dh_{\mathrm{till},s}}{dt}=0$.
Thus $\frac{dw}{dt} = {h_{\mathrm{till},s}} \frac{de}{dt}$, such that
\begin{equation*}
   \frac{de}{dt} = \frac{m}{h_{\mathrm{till},s}}.
\end{equation*}
 As basal sliding velocity $u_b$ is assumed to solely be a result of till deformation, it is determined by the balance between driving stress $\tau_d$, basal frictional stress $\tau_f$, and an unspecified lateral stress which is assumed to be comparatively insignificant. $\tau_d$ is defined as
\begin{equation}\label{eq:tau_d}
\tau_d = \rho_I g \frac{H^2}{L},
\end{equation}
which functions as an approximation of the driving stress with acceleration due to gravity $g$.
As such $u_b >0$ only in the thawed till regime. This is given by
\begin{equation}\label{eq:u_b_t}
u_b =
\frac{A_g W^{n+1}}{4^n (n+1) H^n}
\max\!\left[ \tau_d - \tau_f,\, 0 \right]^n,
\end{equation}
where $A_g$ is Glen's flow law rate factor describing constant creep in the shear margins and $n$ is the Glen flow law exponent.

Basal sliding velocity $u_b$ determines the rate of ice volume lost from sliding, assuming an abrupt end of the glacier where thickness vanishes. This effect is balanced by the rate of accumulation from precipitation $a_c$ which is assumed constant across the ice stream surface:
\begin{equation*}
    \frac{dV}{dt} = a_cLW - W Hu_b.
\end{equation*} 
As the footprint of the ice stream is assumed static, this balance is equivalently described through a change in the thickness of the ice stream, 
\begin{equation}\label{eq:H}
    \frac{dH}{dt} = a_c - \frac{u_bH}{L}.
\end{equation} 

\subsection{Variability in Robel's model}

Two dynamic modes of R13 are identified in \cite{R13}: steady-streaming and binge-purge oscillations, each with two variants dependent on thermomechanical processes. This results in a total of four characteristic behaviours prescribed at different surface temperatures. The steady streaming behaviour is characterised by high geothermal heat flux and warmer surface temperatures, reached by two different pathways dependent on drainage. For relatively warmer temperatures, drainage is always active, sustaining a constant basal velocity. However, when drainage is not sustained, transient oscillations occur. 
 
Lower surface temperatures and geothermal heat flux result in sustained binge-purge oscillations. A weak negative heat flux results in stagnation occurring above the till consolidation threshold, whereas a `stronger' negative heat budget from cooler temperatures enables strong binge-purge oscillations through till freezing and basal cooling, resulting in a longer stagnation and a greater volume of discharge.

\subsection{Kypke's coupled divergent ice streams model} \label{sec:Kypke}

 \begin{figure}[!htbp]
        \centering
        \begin{tikzpicture}
            \node[inner sep=0pt] (img) {\includegraphics[width=\textwidth]{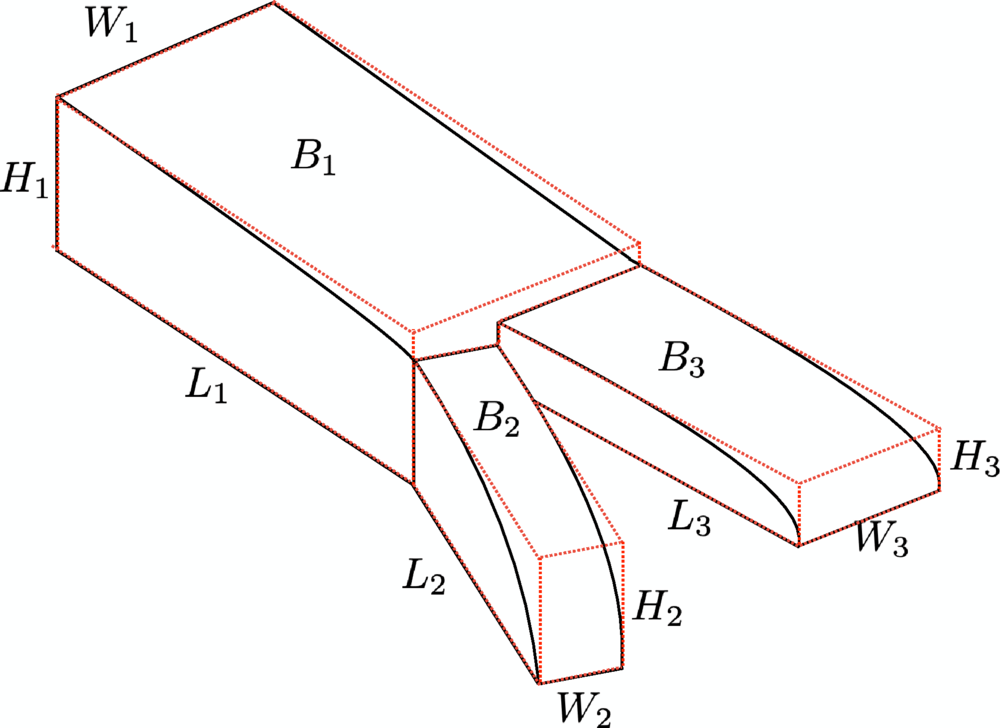}};
            
            \draw[red, line width=2.5pt, -{Stealth[length=10pt, width=8pt]}] (img.west) ++ (-0.2,0.2) -- ++(5.2,-3.6);
        \end{tikzpicture}
        \caption{Box diagram of a divergent topology (taken from \cite{K26}) where $B_1$ gains volume from accumulation and loses volume to $B_2$ and $B_3$ due to streaming flow. $B_2$ and $B_3$ gain volume from both volume flux from $B_1$ and accumulation, and lose volume to streaming flow.}
        \label{fig:div_top}
    \end{figure}

Kypke \textit{et al.} \cite{K26} model a divergent topology of R13 ice streams as shown in Figure \ref{fig:div_top}, using the R13c `per till state' convention. Consequently, K26 has twelve prognostic variables, the same four as R13 for all three boxes, where $i\in\{1,2,3\}$ denotes the corresponding box as illustrated in Figure \ref{fig:div_top}. In this configuration, the driving stress of the upstream box is modulated by the slopes of the two downstream ice streams to account for the independent buttressing effects from the downstream boxes, thus coupling the ice streams: 
\begin{equation}\label{eq:tau_d_1} 
\tau_{d,1}
=
\rho_I g \frac{H_1}{L_1}
\left(
H_1
-
\frac{W_2}{W_1} H_2
-
\frac{W_3}{W_1} H_3
\right).
\end{equation}
Note that the driving stresses of the lower two ice streams continue to use the R13 formulation.

Additionally, volume from the shared reservoir $B_1$ must be distributed to downstream ice streams $B_2$ and $B_3$, thus the thickness of $B_1$ is dependent on the dimensions of its two distributaries. K26 uses the same surface slope modulation as for the driving stress,
\begin{equation}\label{eq:H1_K26}
\frac{d H_1}{d t}
=
a_c
-
\frac{\left(
H_1
-
\frac{W_2}{W_1} H_2
-
\frac{W_3}{W_1} H_3
\right) u_{b,1}}{L_1},
\end{equation}
where $W_1 = W_2 + W_3$. Similarly, the change in the thickness of the downstream distributaries $B_2$ and $B_3$ is dependent on the thickness of $B_1$ such that any volume loss from $B_1$ due to basal sliding enters the distributaries. Additionally, the coupling is unidirectional in maintaining mass balance; the volume of the distributaries can only gain from $B_1$ when the volume of $B_1$ decreases, which is formalised as follows: 
\begin{equation}\label{eq:K26_H2,3}
L_i W_i \frac{d H_i}{d t}
=
\begin{cases}
L_i W_i a_c
-
W_i H_i u_{b,i}
-
\frac{W_i}{W_1}\,\frac{d V_1}{d t},
& \text{if } \dfrac{d V_1}{d t} < 0 \\[1em]
L_i W_i a_c
-
W_i H_i u_{b,i},
& \text{otherwise.}
\end{cases}
\qquad i = 2,3 
\end{equation}
 
Thus for the K26 model, the per-box equations for driving stress $\tau_d$ and ice stream thickness $H$ are modified to allow for both buttressing and mass flux effects respectively. The remaining state equations are unchanged, but are prescribed per box such that the ice stream frequencies evolve independently whilst maintaining a physical coupling. Parametrising each constituent ice stream differently allows for bi-modality, where the system can exhibit both the steady streaming and binge-purge modes depending on the till content of each box.

The model K26 displays all of the same variability as R13, and in addition can admit chaotic behaviour through the interaction of $B_2$ and $B_3$. Various routes to chaotic behaviour occur during the binge-purge mode through period-doubling bifurcations and intermittent chaotic windows. This model also demonstrates hysteresis, bi-stability and strange attractors, such that chaotic transients occur. Transient chaotic behaviour was also demonstrated in \cite{K26b} through the same mechanism; thus the authors surmise this to be a fundamental behaviour of ice streams with a divergent topology.

\section{Correcting the Kypke et al. formulation and results}
\label{sec:correct}

\subsection{Till thickness equation error} 
Comparing equation (\ref{eq:h_till}) (which is derived directly from R13) to R13c and subsequently K26 reveals a missing factor of $1/e_c$. This results in a `sluggish' model where till thickness evolves at a slower rate than in the corrected equation. As the void ratio consolidation threshold $e_c = 0.3$, including the term results in a model that evolves $\times 3.3.$ faster than the equation in R13c and K26. However, this has minimal impact on the dynamics observed in K26, resulting in the same qualitative behaviour at similar parameters.

\subsection{Volume conservation error}

Examining the coupled ice stream thickness equations reveals that mass flux between boxes is not conserved, with unrealistic additional accumulation being added to $B_2$ and $B_3$, as shown in Appendix \ref{appendix:K26_error}.

In order to maintain volume conservation between the boxes, 
we propose the following formulation: 
\begin{equation}\label{eq:H1_fix}
     \frac{dH_1}{dt} 
    =
     a_c - \frac{H_1 u_{b,1}}{L_1}, 
\end{equation}
which is equivalent to equation (\ref{eq:H}) and
\begin{equation}\label{eq:H2,3_fix}
    \frac{dH_i}{dt} 
    =
     a_c - \frac{H_i u_{b,i}}{L_i} + \frac{H_1 u_{b,1}}{L_i}, 
\end{equation}
where $i = 2,3$. A proof of volume conservation for this formulation can be found in Appendix \ref{appendix:div_fix}. Note that this revised formulation preserves unidirectionality by construction, as mass flux cannot be negative.

As such we model a divergent topology using equations (\ref{eq:H1_fix}) and (\ref{eq:H2,3_fix}), in place of equations (\ref{eq:H1_K26}) and (\ref{eq:K26_H2,3}), to couple the volume flux between $B_1$ and the downstream $B_2$ and $B_3$. The remaining equations remain unchanged with each ice stream evolving independently, following its own set of till thickness, void ratio, basal temperature and velocity equations. Driving stress for $B_1$ followsequation (\ref{eq:tau_d_1}) as described in K26, with driving stress for $B_2$ and $B_3$ following equation (\ref{eq:tau_d}). We hereafter refer to this correction as the Divergent Coupling (DC) model.

\subsection{Diagnosing variability in the corrected Divergent Coupling model}\label{sec:diagnosing_variability}
We explore variability in the DC model by repeating the experiments used in K26 with the revised model, allowing the behaviours to be compared. 


By using a per-box diagnostic time series of void ratio $e$, unfrozen till solid thickness $h_{\mathrm{till},s}$, basal sliding velocity $u_b$ and total volume $V_\mathrm{tot}$, we demonstrate the emergence of chaotic variability. 

\begin{figure}[!htbp]
    \centering
    \includegraphics[width=\textwidth]{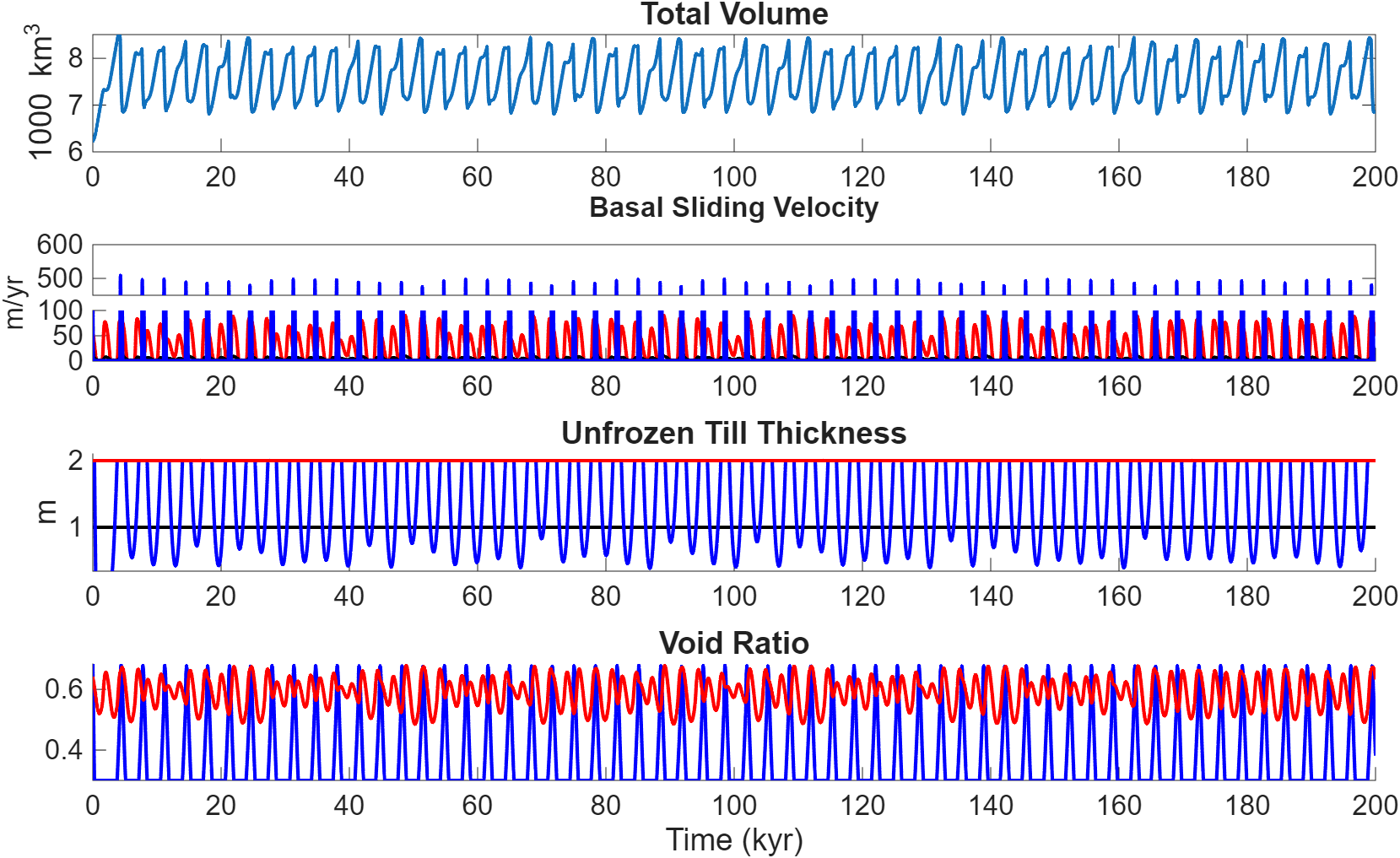}
    \caption{Diagnostic time series of four variables demonstrating chaotic variability in the DC model. $L_1 = 62.8$km, $T_{s,2} = - 13.155^\circ \text{C}$, other parameters as in Table \ref{table:constants}. $B_1$ in black, $B_2$ in blue, $B_3$ in red. The overly large void ratio of $B_1$ is omitted for readability.}
    \label{fig:Chaotic_VolFix_Fig3}
\end{figure}

Figure \ref{fig:Chaotic_VolFix_Fig3} demonstrates each ice stream in a different mode of variability, with $B_1$ in the steady streaming mode (coupling results in volume fluctuations in $B_1$), $B_2$ demonstrating the strong binge-purge mode of variability (due to a partially frozen till), and $B_3$ demonstrating the weak binge-purge mode (as $e_3$ never reaches the consolidation threshold $e_c$). From the highly variable amplitude of $e_3$ and the unrepeated peaks observed in $V_\mathrm{tot}$, we can diagnose this parameter regime as demonstrating chaotic behaviour.


By studying how the number and location of the peaks of total ice stream volume change as the surface temperature of box 2 $T_{s,2}$ is varied, we create bifurcation diagrams demonstrating the routes to chaos in the DC model.

\begin{figure}[!htbp]
    \centering
    \begin{tikzpicture}[
        zoombox/.style={draw=black, thick, inner sep=0pt},
        zoomline/.style={-Stealth, black, thick, dashed},
        caplabel/.style={font=\small, align=center, text width=0.45\linewidth}
    ]

        \node[inner sep=0pt] (main) {
            \includegraphics[width=\textwidth]{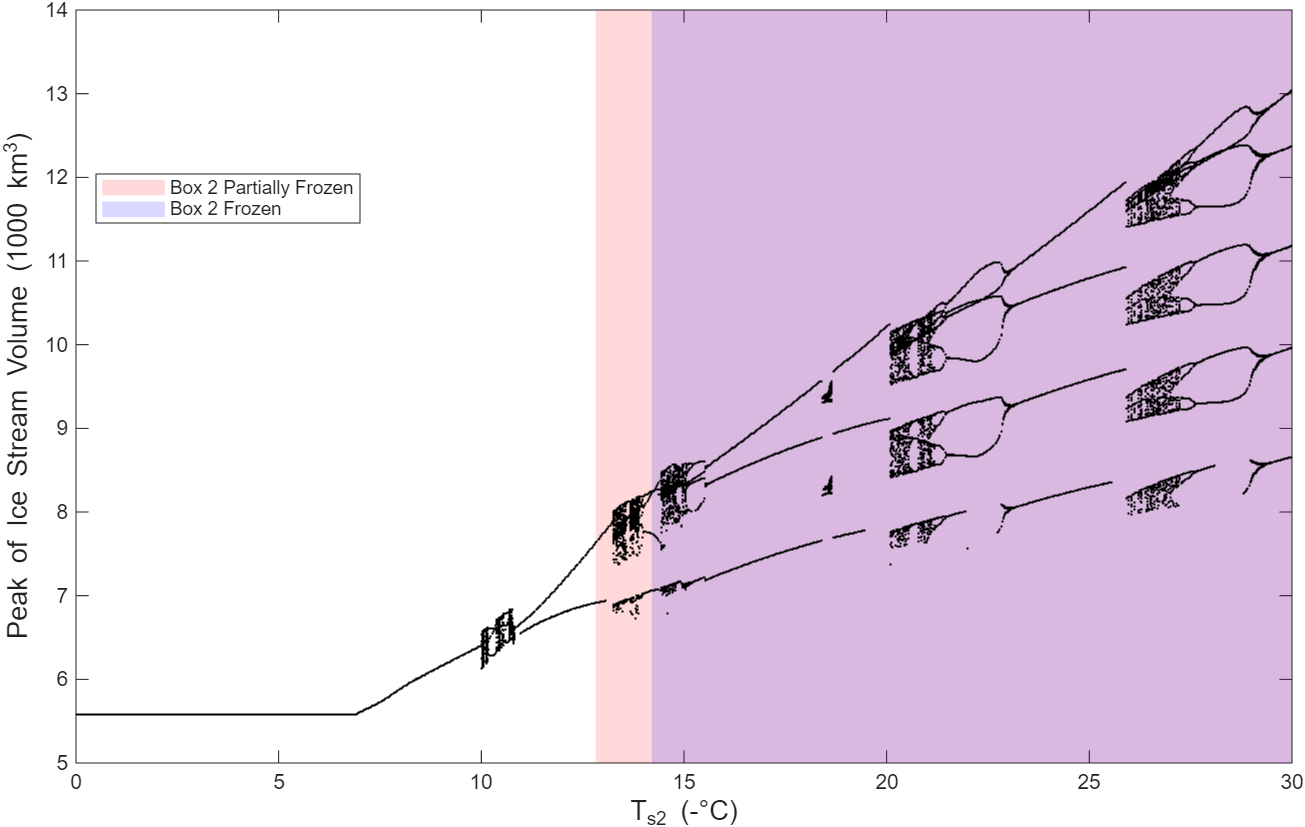}
        };

        \node[zoombox, minimum width=0.6cm, minimum height=1.0cm] (box_b) at (-1.9, -2.8) {};
        
        \node[zoombox, minimum width=1.6cm, minimum height=2.2cm] (box_c) at (0, -1.6) {};

        \node[inner sep=0pt, below=2.5cm of main.south west, anchor=north west] (zoom_b) {
            \includegraphics[width=0.49\linewidth, height=5cm]{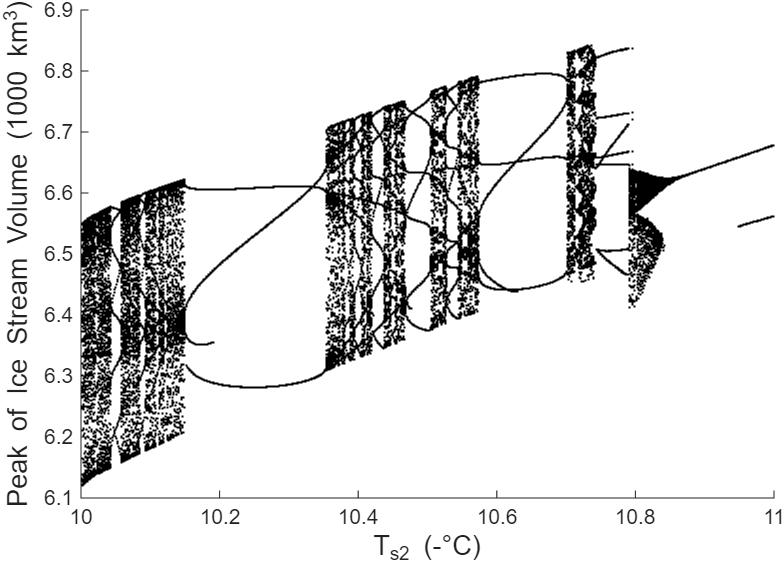}
        };

        \node[inner sep=0pt, below=2.5cm of main.south east, anchor=north east] (zoom_c) {
            \includegraphics[width=0.49\linewidth,height=5cm]{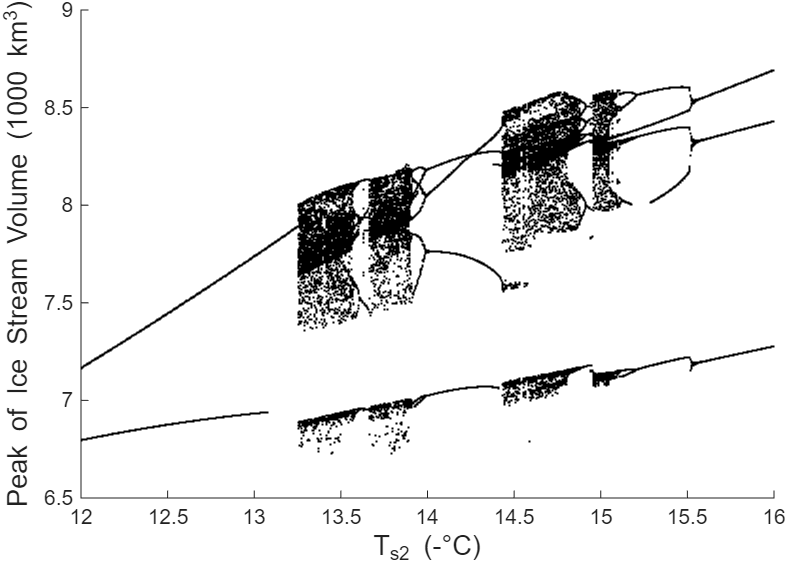}
        };

        \draw[zoomline] (box_b.south) -- (zoom_b.north);

        \draw[zoomline] (box_c.south) -- (zoom_c.north);

        \node[caplabel, below=5pt of main] (cap_a) {
            (a)
        };
        
        \node[caplabel, below=5pt of zoom_b] (cap_b) {
            (b) 
        };

        \node[caplabel, below=5pt of zoom_c] (cap_c) {
            (c)
        };

    \end{tikzpicture}

    \caption{Bifurcation diagrams for the DC model created by varying $T_{s,2}$, other parameters as in Table \ref{table:constants}. The main overview clearly demonstrates a period adding bifurcation with quasiperiodic and chaotic windows in between. (a) Full temperature bifurcation sweep ranging from $T_{s,2} = 0^\circ$C to $-30^\circ$C; (b) Enlarged frequency locking behaviour; (c) Strange attractor emergence.}
    \label{fig:bifurcation_diagrams}
\end{figure}
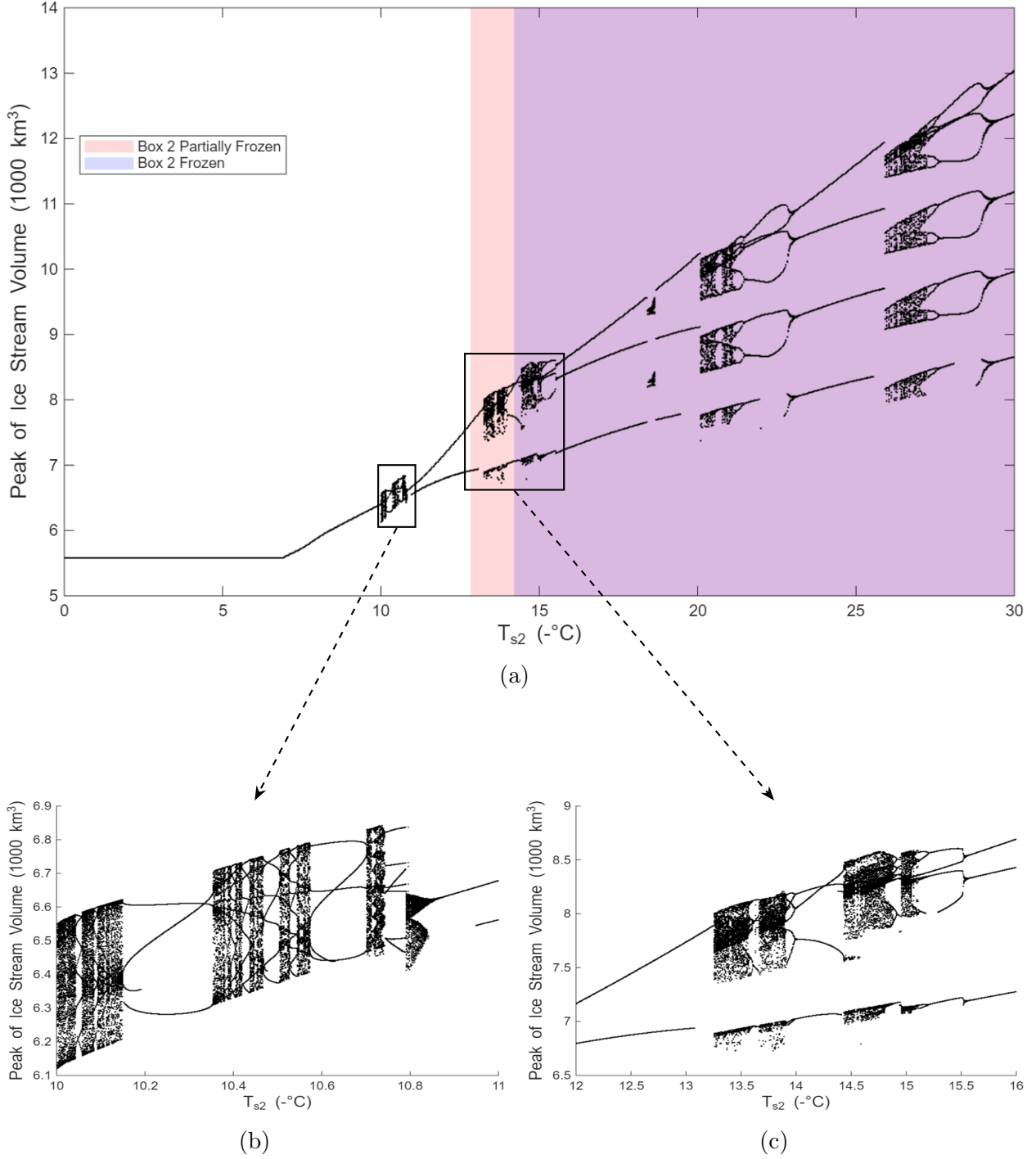

We vary $T_{s,2}$ to visualise how chaos emerges in Figure \ref{fig:bifurcation_diagrams}a. Starting from $0^\circ$C, $T_{s,2}$ begins in the steady streaming mode before entering the oscillating steady streaming mode near $-7^\circ$C, characterised by the increase in the magnitude of volume peaks. Near $-10^\circ$C the system undergoes a Hopf bifurcation, transitioning into the weak binge-purge oscillatory mode through a limit cycle. This results in phase-locking behaviour as shown in Figure \ref{fig:bifurcation_diagrams}b where the period of the orbit increases rapidly, entering quasi-periodic regimes before decreasing again into lower period orbits. This phase-locking behaviour is resolved through a period-adding bifurcation, resulting in a periodic region of period-2. The system undergoes another boundary crisis near $-13^\circ$C resulting in intermittent chaotic behaviour, illustrated by Figure \ref{fig:bifurcation_diagrams}c. This is then resolved into a period-3 orbit, continuing the period-adding cascade.

\begin{figure}[!htbp]
    \centering
    \includegraphics[width=\textwidth]{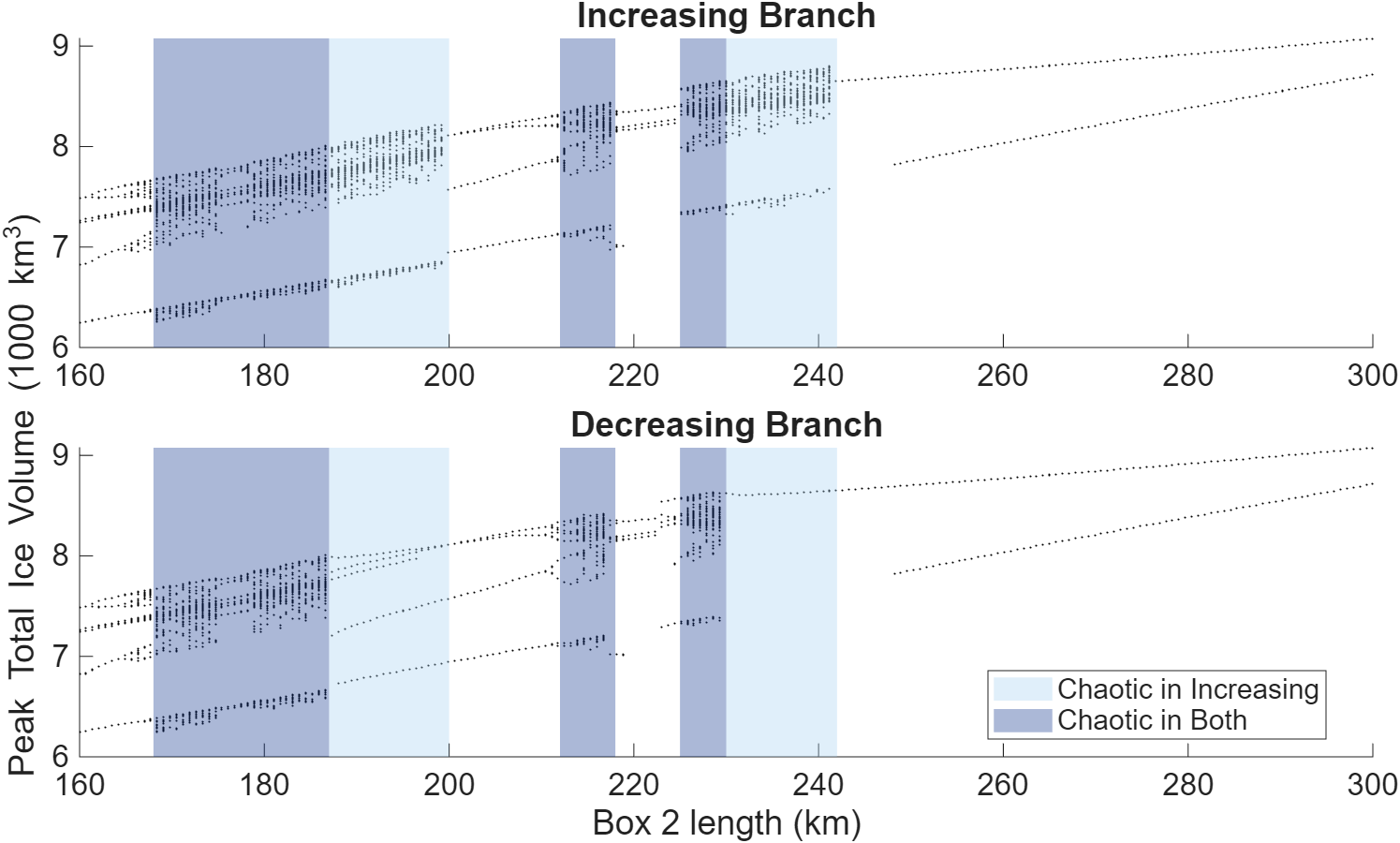}
    \caption{Bifurcation diagrams for the DC model made by increasing and decreasing $L_2$ from $160$km to $300$km, with a fixed $L_1 = 55\text{km}$ which demonstrates hysteresis in two windows. Other parameters as in Table \ref{table:constants}.}
    \label{fig:hysteresis}
\end{figure}

Bifurcation diagrams constructed by varying $L_2$ reveal bi-stability in the DC model, where chaotic behaviour is present in increasing windows but absent in decreasing windows. Figure \ref{fig:hysteresis} demonstrates where chaotic windows occur in both directions, with bi-stability occurring in two windows from  $L_2 = 187$km to $200$km and from $230$km to $242$km.


As in \cite{K26}, we create Poincaré sections by recording when the oscillating box 2 till void ratio $e_2$ crosses the threshold $0.6$ to create a consistent discretisation. This creates a series of 11D recorded model states in the 12D system. We project these model states onto the 2D surface of box 3 void ratio $e_3$ and total ice stream volume $V_{\mathrm{tot}}$ to reveal the periodicity of the trajectory and the shape of the attractor, allowing for the identification of strange attractors.

\begin{figure}[!htbp]
    \centering
    \begin{subfigure}[c]{0.49\textwidth}
        \centering
        \includegraphics[width=\textwidth]{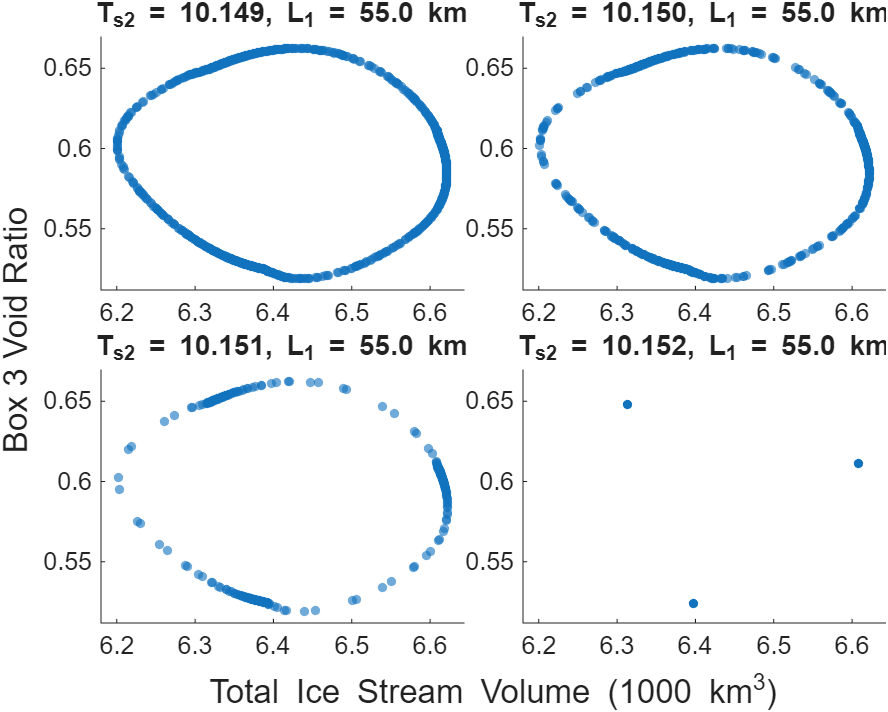}
        \caption{}
        \label{fig:PC_rep_10}
    \end{subfigure}
    \hfill
    \begin{subfigure}[c]{0.49\textwidth}
        \centering
        \includegraphics[width=\textwidth]{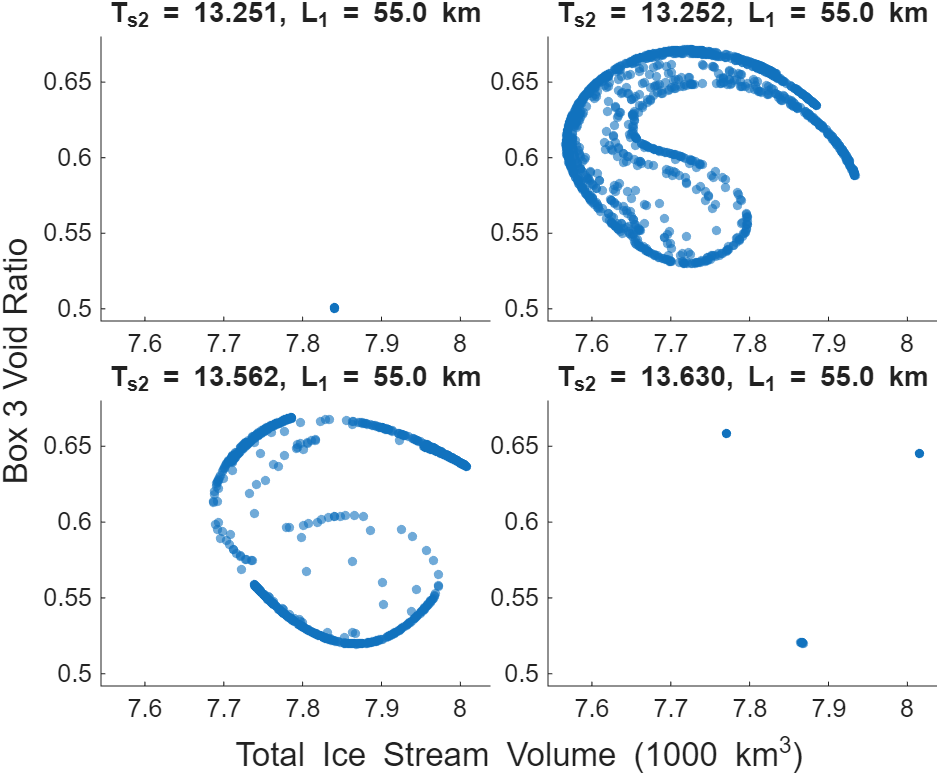}
        \caption{}
        \label{fig:PC_rep_13}
    \end{subfigure}

    \caption{Poincaré sections illustrating different phase space structures when varying $T_{s,2}$ in the DC model, other parameters as in Table \ref{table:constants}. (a) Phase locking where quasiperiodicity resolves to periodicity; (b) The emergence and collapse of a strange attractor.}
    \label{fig:PC_rep_combined}
\end{figure}

The Poincaré sections in Figure \ref{fig:PC_rep_combined} confirm the behaviour observed in both the phase locking and chaotic windows. Figure \ref{fig:PC_rep_13} shows four Poincaré sections (2D intersections of the full phase space, see Section \ref{sec:diagnosing_variability}), demonstrating a boundary crisis occurring at $-13.251^\circ$C which eventually collapses into a period-3 orbit. The circular appearance of these Poincaré sections shows the model undergoes frequency locking through the torus breakdown route to chaos \citep{aronson_bifurcations_1982}. 

\begin{figure}[!htbp]
    \centering
    \begin{subfigure}[b]{0.95\textwidth}
        \centering
        \includegraphics[width=\textwidth]{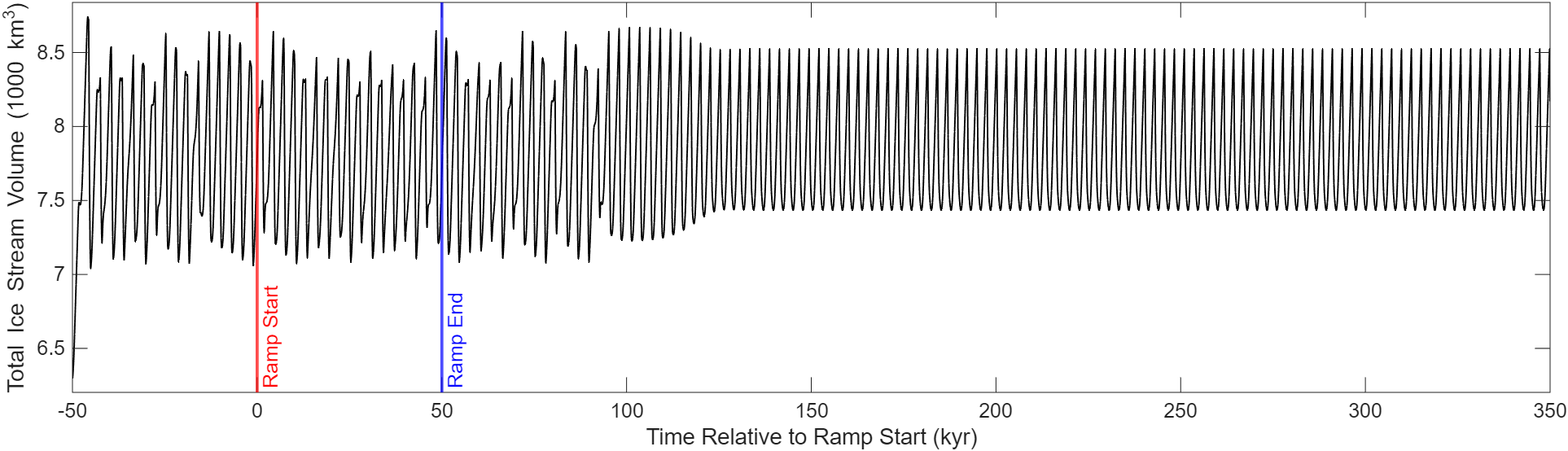}
        \caption{}
        \label{fig:L2_Transient}
    \end{subfigure}

    \vspace{0.4cm} 

    \begin{subfigure}[b]{0.95\textwidth}
        \centering
        \includegraphics[width=\textwidth]{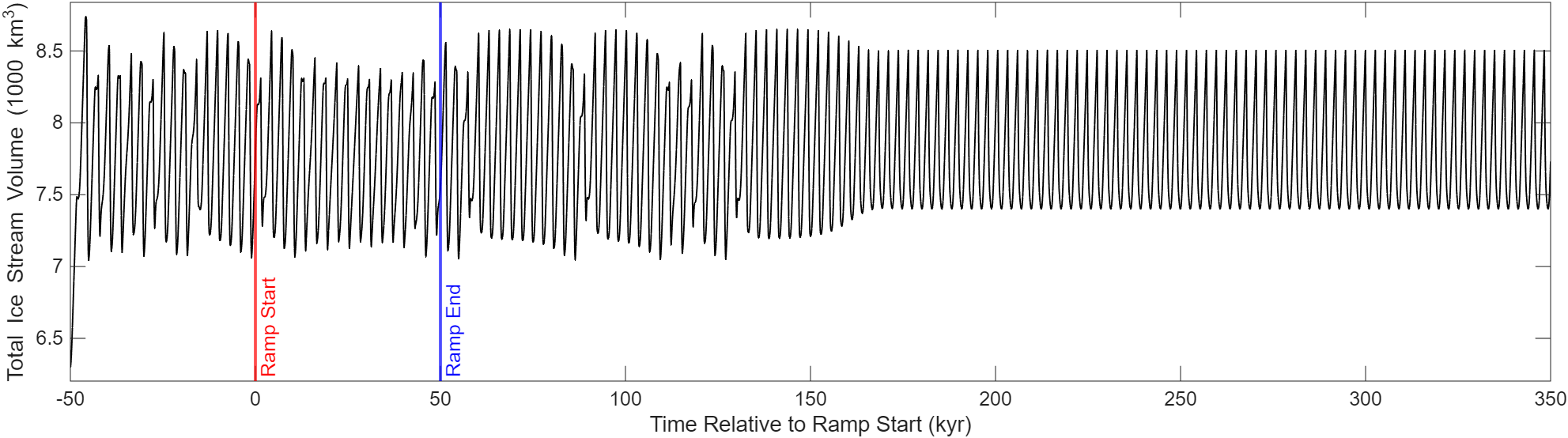}
        \caption{}
        \label{fig:L3_Transient}
    \end{subfigure}

    \caption{Time series of chaotic transients in the DC model where ice stream length is varied over a period of $50$kyr. (a)  $L_2$ is increased by $1$km from $243.2$km to $244.2$km; (b)  $L_3$ is decreased by $1$km from $250$km to $249$km. Other parameters as in Table \ref{table:constants}.}
    \label{fig:L2_L3_Transient_Combined}
\end{figure}

By replicating the ramping test performed on K26 in \cite{K26}, we confirm that chaotic transients occur in the DC model. The time series in Figure \ref{fig:L2_Transient} show initialising the model on a chaotic trajectory for $50$kyr before increasing $L_2$ by $1$km over another $50$kyr period such that the model is in a periodic region. However, the time series shows chaotic behaviour continues for over $40$kyr after the ramp ends, thus demonstrating a chaotic transient. Similarly, decreasing $L_3$ by $1$km results in a similar (but longer) chaotic transient (Figure \ref{fig:L3_Transient}).

\subsection{Impact of corrections to the Kypke et al. model}

Model behaviour is notably impacted by correcting volume conservation in K26 to create the DC model. We observe both quantitative impacts such as the removal of additional accumulation decreasing the magnitude of binge-purge events, as well as qualitative impacts. K26 presented both a period-doubling bifurcation and intermittent chaos occurring through a series of saddle-node bifurcations \citep{PM}. While intermittent behaviour is retained in the DC model, we observe a period-adding cascade route to chaos. This is a consequence of border-collision bifurcations, which directly result from the interaction of orbits with non-smooth boundaries \citep{di_bernardo_piecewise-smooth_2008}. In continuous piecewise-smooth systems such as this one, local dynamics from border collisions reduce to a circle map, producing the period-adding cascade and consequent phase locking \citep{di_bernardo_piecewise-smooth_2008, granados2015}.

\section{A two-parameter study of variability in the Divergent Coupling model}\label{sec:sync_chaos}

Phase locking in the Arnold circle map results in both synchronisation and chaos, depending on the specific coupling strength and frequency detuning parameters. As such, we identify candidate parameters to vary in order to demonstrate this behaviour before quantifying the chaotic behaviour observed.  
\subsection{Frequency locking synchronisation} 
To demonstrate frequency locking behaviour and Arnold tongues in the DC model, we require prescribed constants that can be varied per box which act as frequency detuning and coupling strength variables. We find $T_{s,2}$ acts as a frequency detuning variable, and $1/L_1$ acts as a coupling strength variable as increasing $L_1$ decreases $\tau_{d,1}$, which in turn reduces $u_{b,1}$ and thus the mass flux entering $B_2$ and $B_3$.

Frequency locking is evaluated through a rotation number, which we determine by approximating the raw frequencies of oscillators $B_2$ and $B_3$ to be the frequencies of their till void ratios $e_2$ and $e_3$. These highly variable parameters are only directly dependent on the individual state of each box, making them strong diagnostic tools for comparing behaviour between the two downstream ice streams. 
We calculate the frequency of $e_2,e_3$ by recording the number and time of peaks in each void ratio time series. The number of peaks is equivalent to the number of complete cycles, thus we obtain the average frequency across the time series: 

\begin{equation*}
    f_i = \frac{\text{number of peaks}-1}{\text{time of last peak}-\text{time of first peak}}
=\frac{\text{number of cycles}}{\text{total time elapsed}}.
\end{equation*}
As such we approximate the rotation number $\rho$ by taking the ratio of the two frequencies ($f_2,f_3$) modulo 1 to obtain the relative phase of $B_2$ against $B_3$,\\
\begin{equation*}
    \rho = \frac{f_2}{f_3} \mod{1}.
\end{equation*}
When the rotation number is rational such that $\rho = p/q \in \mathbb{Q}$, we consider the system to be exhibiting frequency locking. 

\begin{figure}[!htbp]
    \centering
    \includegraphics[width=\textwidth]{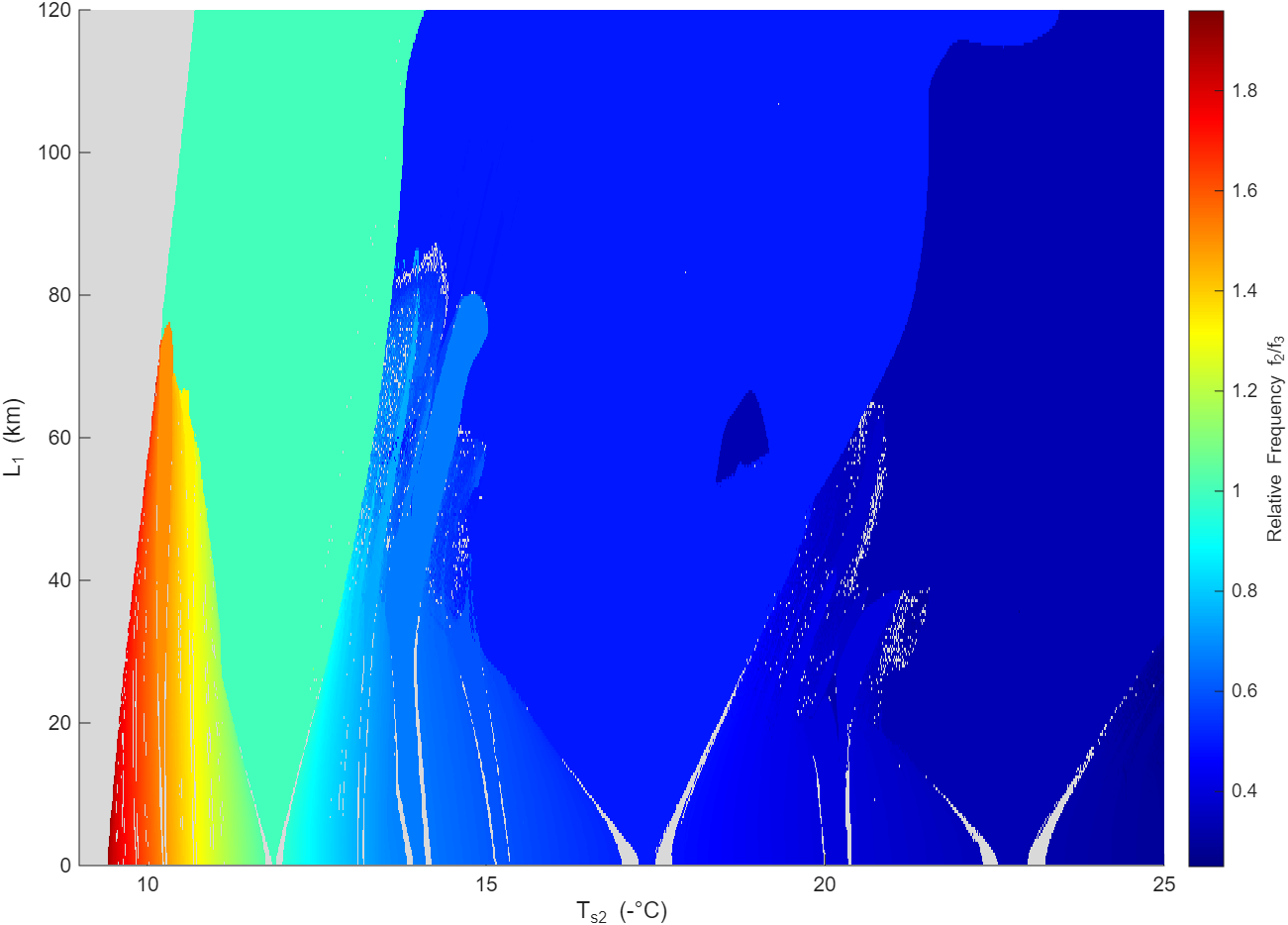}
    \caption{Parameter sweep varying $L_1$ and $T_{s,2}$, demonstrating frequency locking $f_2/f_3$ in the DC model where the maximum denominator of the rotation number $q = 32$ such that the main ‘tongues’ are apparent.}
    \label{fig:freq_rat}
\end{figure}

Figure \ref{fig:freq_rat} illustrates this, calculating the rationality of the frequency ratio with a maximum denominator of $q =32$ to identify the main `tongues' (areas of phase locking). Below $L_1 = 100$km strong coupling results in synchronisation and phase locking such that Arnold tongues are apparent. The grey triangular area on the left of these plots represents the steady streaming mode for $B_2$, thus no oscillations are recorded. As $T_{s,2} $ approaches $ -9.5^\circ$C we observe 2:1 frequency locking between $B_2$ and $B_3$ which decreases to 1:1 locking at $T_{s,2} \approx -12^\circ$C. Frequency locking is similarly varied until 2:3 locking occurs for $T_{s,2} \approx -13.5^\circ$C. After this as $T_{s,2} \rightarrow -25^\circ$C, the rate of frequency change is much slower; however, phase-locking regions are still apparent. Between these regions of Arnold tongue-like behaviour exist large regions of phase locking which decrease in phase from 1:1 locking to 1:2 and then 1:3 locking as $T_{s,2}$ decreases.

\begin{figure}[!htbp]
    \centering
    \begin{subfigure}[b]{0.51\textwidth}
        \centering
         \includegraphics[width=\textwidth, height = 5.1cm]{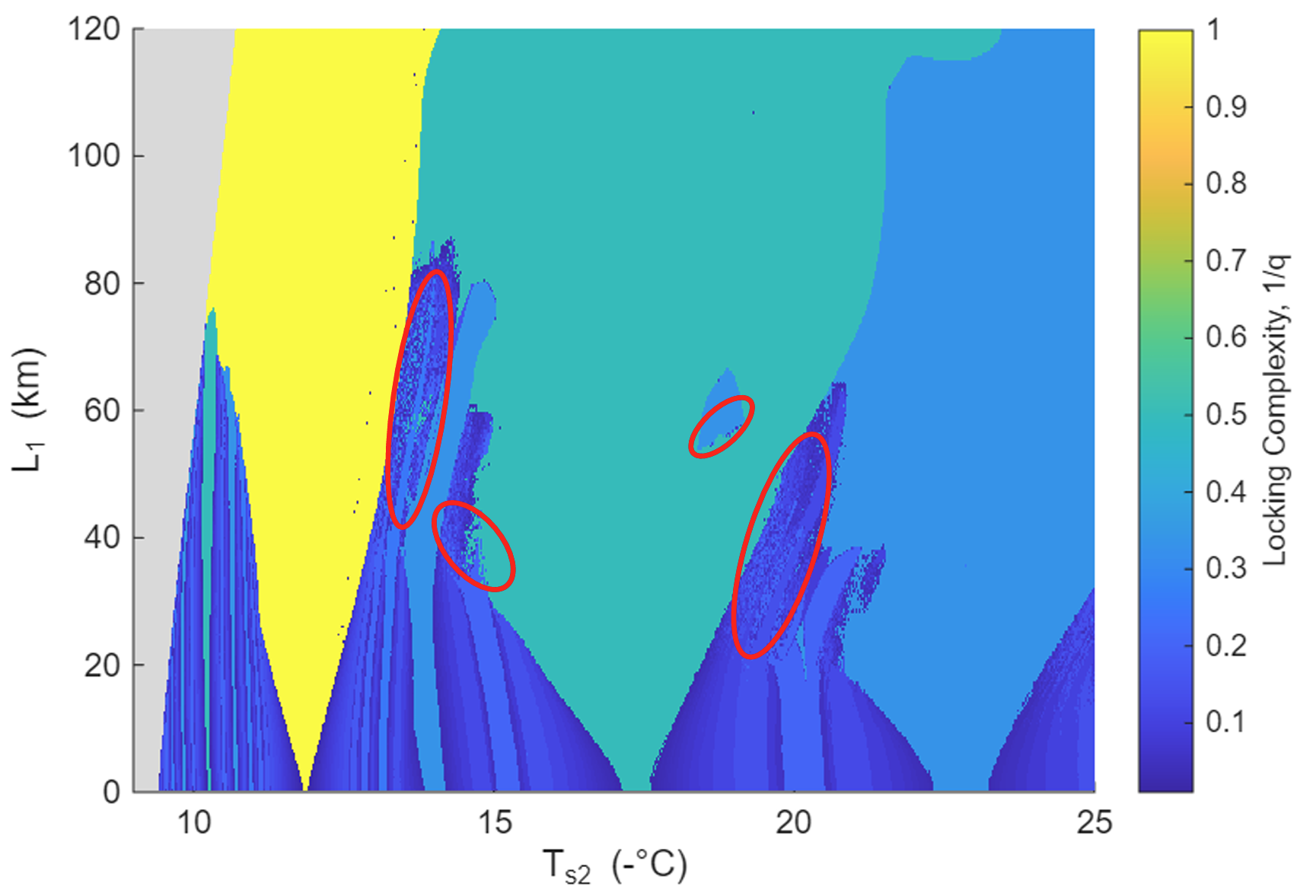}
        \caption{}
        \label{fig:complex}
    \end{subfigure}
    \hfill
    \begin{subfigure}[b]{0.47\textwidth}
        \centering
        \includegraphics[width=\textwidth, height = 5.0cm]{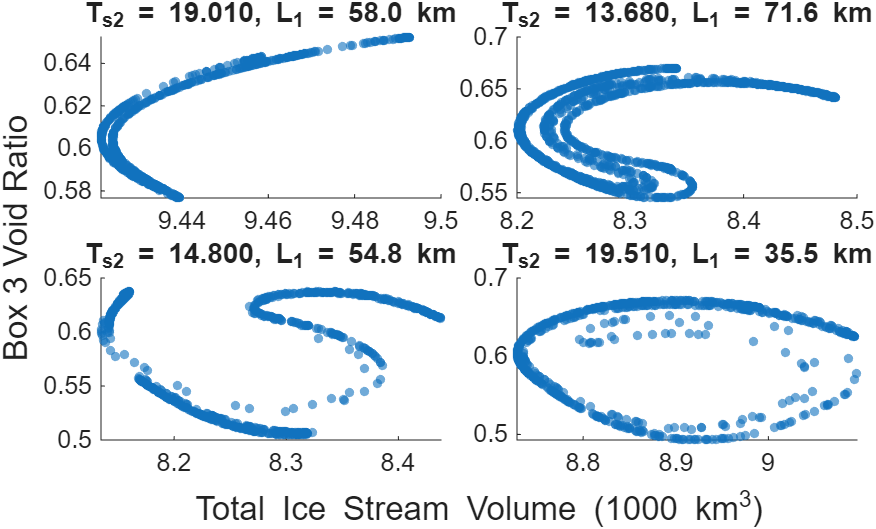}
        \caption{}
        \label{fig:PC_chaos}
    \end{subfigure}

    \caption{Variability in the DC model is highlighted in different parametrisations. (a) Parameter sweep  (as in Figure \ref{fig:freq_rat}) of the inverse of the denominator of the rotation number $1/q$, with chaotic regions highlighted by red ellipses; (b) Poincaré sections demonstrating the different behaviours in the chaotic regions.}
    
    \label{fig:chaos_and_poincare_combined}
\end{figure}

However, this does not directly distinguish high-period phase locking from dense quasiperiodicity. Thus we identify quasiperiodic behaviour through considering $1/q$. 
For quasiperiodic regimes the rotation number is irrational ($\rho \notin \mathbb{Q}$), such that trajectories densely cover the invariant torus without strictly closing, resulting in $q \to \infty$. As such when $1/q \ll1$ the total period of the system has a sufficiently large periodic orbit such that its period is indistinguishable from an irrational rotation number, such that we identify quasiperiodicity. This is shown in Figure \ref{fig:complex} which demonstrates clear tongues with dark blue areas of quasiperiodicity.

 As the 2D frequency locking sweeps are unable to discern between quasiperiodic and chaotic behaviour, Poincaré sections in Figure \ref{fig:PC_chaos} illustrate the phase space in highly variable regions circled in Figure \ref{fig:complex}.  

\subsection{Quantifying chaotic behaviour} 

\begin{figure}[!htbp]
    \centering
    \begin{subfigure}[b]{0.48\textwidth}
        \centering
        \includegraphics[width=\textwidth, height = 4.5cm]{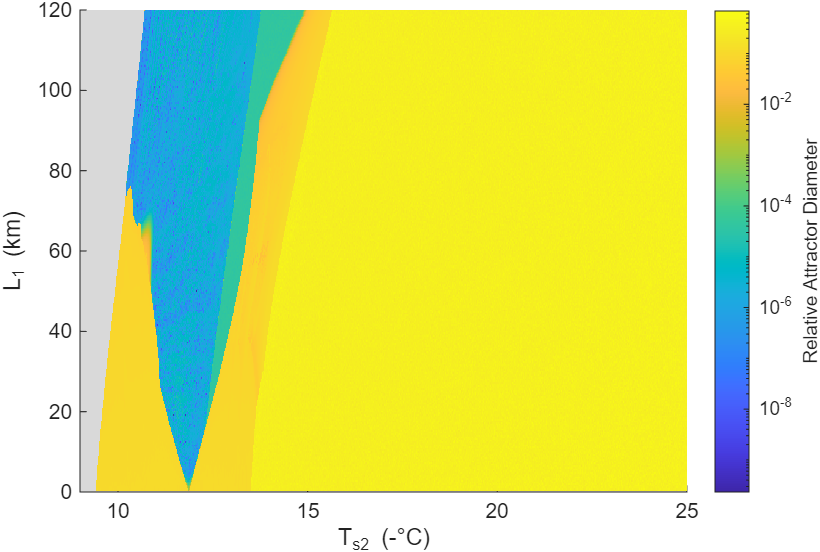}
        \caption{}
        \label{fig:att_diam}
    \end{subfigure}
    \hfill
    \begin{subfigure}[b]{0.48\textwidth}
        \centering
        \includegraphics[width=\textwidth, height = 4.5cm]{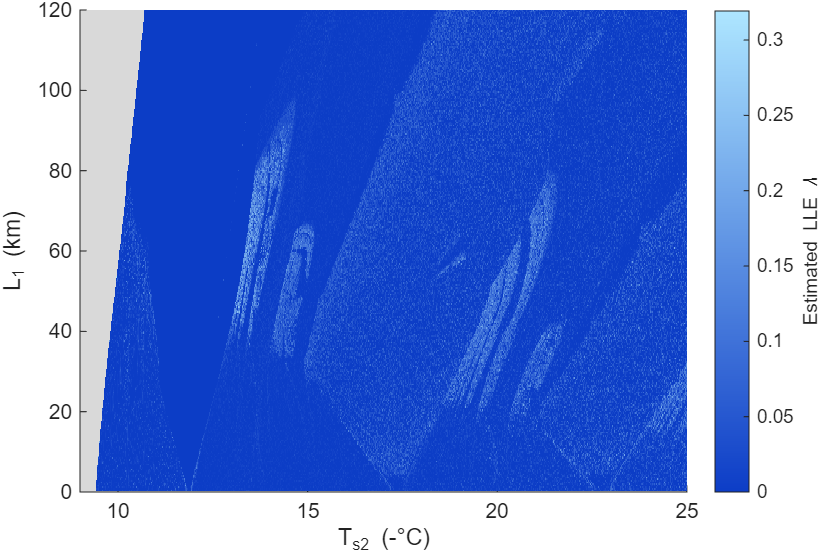}
        \caption{}
        \label{fig:LLE}
    \end{subfigure}

    \vspace{0.00cm} 

    \begin{subfigure}[b]{0.48\textwidth}
        \centering
        \includegraphics[width=\textwidth, height = 4.5cm]{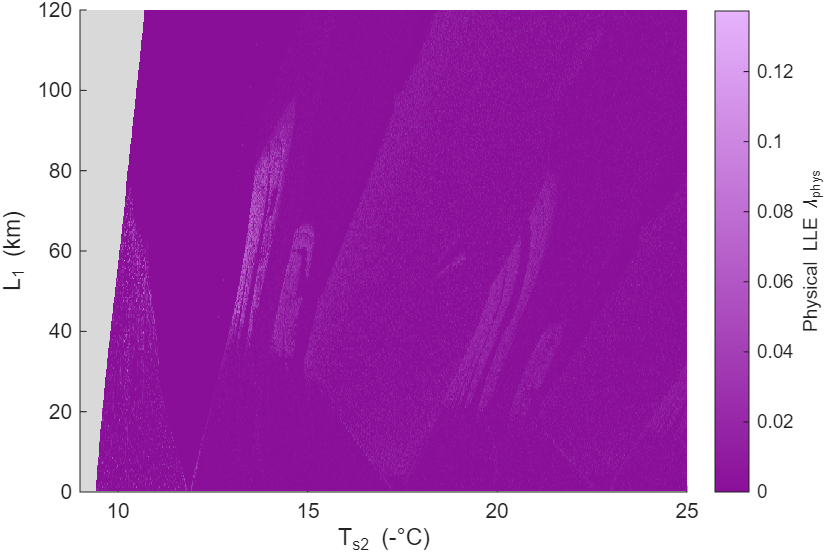}
        \caption{}
        \label{fig:Phys_LLE}
    \end{subfigure}
    \hfill
    \begin{subfigure}[b]{0.48\textwidth}
        \centering
        \includegraphics[width=\textwidth, height = 4.5cm]{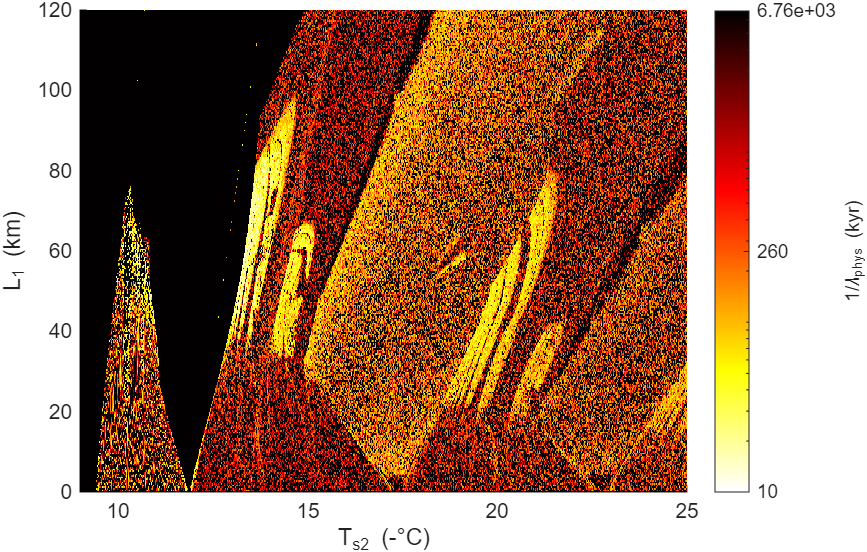}
        \caption{}
        \label{fig:efold_time}
    \end{subfigure}
\vspace{-0.3cm} 
    \caption{Diagnostic plots for chaos in the DC model created using Rosenstein's algorithm adapted for Poincaré sections. The parameter sweep is the same as Figure \ref{fig:freq_rat}. (a) Attractor diameter; (b) LLE estimated per binge-purge oscillation cycle $\lambda$; (c) Physical LLEs, where the LLE timescale is standardised to physical system time ($1/\mathrm{kyr}$) ($\lambda_\mathrm{phys}$); (d) Timescale ($1/\lambda_\mathrm{phys}$).}
    \label{fig:lyapunov_diagnostics_grid}
\end{figure}

To quantify predictability in the chaotic regimes, we estimate the Largest Lyapunov Exponent (LLE) by creating $11$D Poincaré sections from $12$D phase space, saving the full system as $e_2$ increases through $0.6$, and comparing successive sections to track the divergence of nearby trajectories using an adapted form of Rosenstein's Algorithm (see Appendix \ref{appendix:Rosn}). Through this method we also obtain the normalised attractor diameter across the 2D parameter sweep, which is used both to mask period-1 behaviour as specifically non-chaotic, and as a secondary diagnostic to understand underlying behaviour. This is illustrated in Figure \ref{fig:att_diam}, which clearly shows where period-1 dynamics end and where higher period dynamics (and the potential for chaotic regimes) begin. This is applied in Figure \ref{fig:LLE} where period-1 behaviour is masked with an LLE of $\lambda = 0$ to remove false positives. Note that we only present $\lambda \ge 0$ as the algorithm is only accurate for divergent behaviour. Thus darker areas are likely periodic, with progressively lighter shading corresponding to increasingly chaotic behaviour.

As the phase space is recorded per cycle, average cycle length becomes the natural unit of divergence for the LLEs. As the average length of a cycle varies across the parameter sweep, we calculate physical LLEs $\lambda_\text{phys}$, which have units of ${1}/{\text{seconds}}$, by dividing $\lambda$ by the respective average period. Figure \ref{fig:Phys_LLE} illustrates $\lambda_\mathrm{phys}$ and follows the same shading convention as Figure \ref{fig:LLE}, demonstrating likely chaotic behaviour with a more accurate relative rate of trajectory divergence. Converting this rate of physical divergence into a timescale by taking $1/\lambda_\mathrm{phys}$, we obtain Figure \ref{fig:efold_time}. This clearly demonstrates that chaotic regimes correspond to notably shorter timescales, where nearby trajectories diverge exponentially on faster timescales. Note that the chaotic regions identified in Figure \ref{fig:LLE} correspond to the `grainy' regions highlighted in Figure \ref{fig:complex}, as the period of these regions is highly sensitive to small differences in initial conditions.

\subsection{Switches in the non-linear dynamics}

\begin{figure}[!htbp]
    \centering
    \begin{subfigure}[b]{0.48\textwidth}
        \centering
        \includegraphics[width=\textwidth, trim=0 0 0 3.5mm, clip]{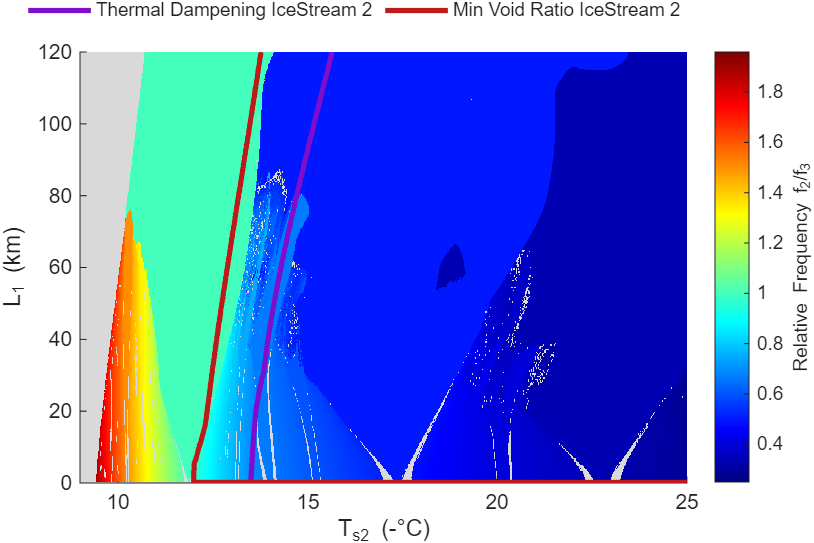}
        \caption{}
        \label{fig:switching_freq}
    \end{subfigure}
    \hfill
    \begin{subfigure}[b]{0.48\textwidth}
        \centering
        \includegraphics[width=\textwidth, trim=0 0 0 3.5mm, clip]{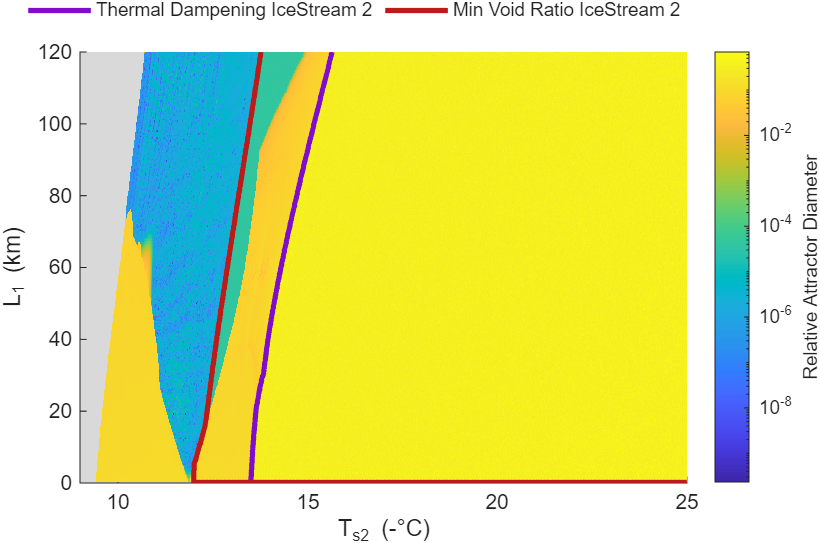}
        \caption{}
        \label{fig:switching_cloud}
    \end{subfigure}
    \vspace{-0.2cm} 
    \caption{Non-linear switches in the DC model when equations (\ref{eq:h_till}) and (\ref{eq:Tb}) activate on crossing the red and purple lines respectively, overlaying parameter sweeps of: (a) Frequency locking (Figure \ref{fig:freq_rat}); (b) Attractor diameter (Figure \ref{fig:att_diam}). Crossing a boundary from left to right implies the switch is activated. }
    \label{fig:switching_diagnostics_combined}
\end{figure}

In Figure \ref{fig:switching_diagnostics_combined}, switches in the non-linear dynamics when till changes state are overlaid on Figures \ref{fig:freq_rat} and \ref{fig:complex} to illustrate where the model enters the partially frozen till state (activating equation \ref{eq:h_till}), which triggers the strong binge-purge mode, and the frozen till state (activating equation \ref{eq:Tb}) respectively. Note that only the mode of variability of $B_2$ changes; thus no crossing occurs in $B_1$ or $B_3$. It is clear that no chaotic behaviour is observed before the strong binge-purge mode is activated (when the red line is crossed from left to right). Furthermore, the frozen switch in purple is aligned with the greater attractor diameter on the right of Figure \ref{fig:switching_cloud}.

\section{Consequences for variability in coupled ice streams}
\label{sec:consequences}

\subsection{Variability in the Divergent Coupling model}

Synchronisation and chaos arise in the DC model from the competition between the different frequencies imposed on $B_2$ by $T_{s,2}$, and coupling strength $1/L_1$, which modulates the relationship between $B_2$ and $B_3$. For a sufficiently strong coupling, the oscillators synchronise such that the attractor is on a torus and Arnold tongues emerge, decreasing in width as $L_1$ increases. For a weaker coupling strength, the natural frequencies of $B_2$ and $B_3$ compete with the coupling; thus the torus is lost \citep{vadivasova_synchronization_1999}. 

When $B_2$ is in the weak binge-purge mode, both ice streams are in the same mode where no border collision occurs, thus the torus reduces to a period-1 orbit smoothly. When $B_2$ is in the strong binge-purge mode, border collisions occur as coupling strength decreases, resulting in strange attractors from a chaotic torus breakdown \citep{di_bernardo_piecewise-smooth_2008} (see Figure \ref{fig:chaos_and_poincare_combined}). 
Figure \ref{fig:switching_freq} supports this mechanism, where crossing the switch to strong binge-purge oscillations (red) admits chaos. The second switch admits a frozen till (purple) such that before the switch, decreasing  $T_{s,2}$ lengthens stagnation directly and the locking ratio varies rapidly in response. Once the till is frozen, further cooling only decreases basal temperatures below the melt threshold $T_b > T_m$, leaving till state unchanged. As such the dependence of stagnation duration on $T_{s,2}$ is weaker and the locking ratio changes slowly in response. This is reflected in the attractor diameter which increases with each switch and is constant after the second (Figure \ref{fig:att_diam}), indicating that the mechanism generating chaotic variability is fixed once the frozen regime begins.

\subsection{Implications of different types of variability on ice stream predictability}

The minimum timescale in the chaotic regions shown in Figure \ref{fig:efold_time} is $10$kyr, suggesting that even in the most chaotic regions behaviour remains largely predictable on short timescales. However, the presence of synchronisation in the dense Arnold tongues demonstrated in Figure \ref{fig:freq_rat} implies a sensitivity to trajectories in non-chaotic regions which is consequently invariant to timescale.

Chaotic transients may also occur regardless of timescale; however, these are rare in the parameter space, confined to the chaotic regions illustrated in Figure \ref{fig:LLE}. This suggests that the behaviour observed in \cite{K26b}, where chaotic transients in a retreating ice stream prevent the tipping of the Greenland Ice Sheet, is highly dependent on model parametrisation. Demonstrating that chaotic transients in the DC model can occur for both retreating and advancing ice streams provides further evidence of a mechanistic origin to this behaviour, compared to K26, which only demonstrates a chaotic transient from ice stream advance.

However, variable behaviour in retreating ice streams is not guaranteed to be chaotic. As the model is scale invariant (assuming the shallow shelf assumption holds \citep{macayeal_bingepurge_1993}), increasing $L_1$ has an equivalent effect to simultaneously decreasing $L_2$ and $L_3$ proportionally. Thus a retreating pair of ice streams with a divergent topology may pass through an Arnold tongue in parameter space, exhibiting highly variable behaviour. 

On the timescale of Heinrich events and Dansgaard-Oeschger events, whether an ice stream exists in a synchronous or chaotic state greatly impacts variability.
Mann \textit{et al.} \cite{MannRobel} demonstrate how an R13 ice stream model coupled to an ocean box model results in similar frequency-locking (Arnold tongue) behaviour, suggesting a plausible mechanism for the irregular timing between Heinrich events and Dansgaard-Oeschger events. Examining the behaviour of the DC model when in a chaotic regime shows the internal binge-purge oscillations take $\sim3\,\text{kyr}$, while the minimum timescale of divergence from the adjusted LLE is $\sim 10\,\text{kyr}$, which translates to an error-doubling time of $\sim 7\,\text{kyr}$. This timescale, similar to the timescale of Heinrich events \citep{macayeal_bingepurge_1993}, suggests that variability of ice streams in a divergent topology may limit the predictability of paleoclimate models.

\subsection{Comparison to a model of ice streams with convergent coupling}

\begin{figure}
        \centering
        \begin{tikzpicture}
            \node[inner sep=0pt] (img) {\includegraphics[width=\textwidth]{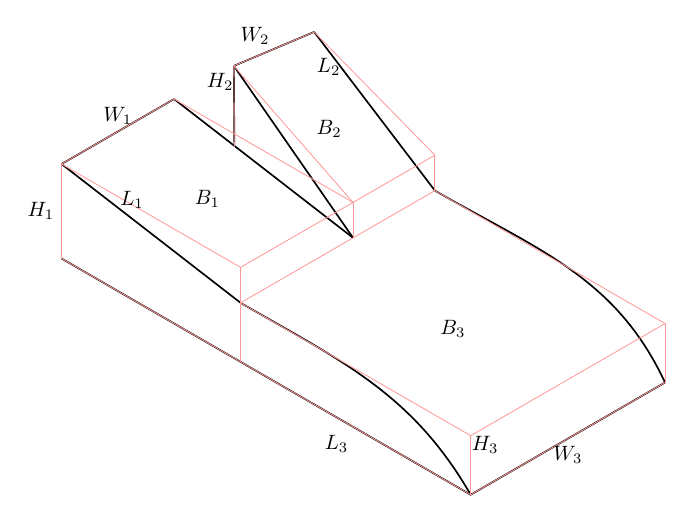}};
            
            \draw[red, line width=2.5pt, -{Stealth[length=10pt, width=8pt]}] (img.west) ++ (-0.2,0.2) -- ++(5.2,-3.6);
        \end{tikzpicture}
        \caption{ Box diagram of a convergent topology where $B_1$ and $B_2$ gain volume from accumulation and lose volume to $B_3$ due to streaming flow. $B_3$ gains volume due to volume flux from $B_1$ and $B_2$ and from accumulation, and loses volume due to streaming flow.}
        \label{fig:conv_top}

 \end{figure}

A convergent topology can be similarly modelled by considering two upstream ice streams ($B_1$ and $B_2$), where both their volume fluxes enter a shared terminus ($B_3$) which in turn exerts a driving stress on the upstream ice streams, as illustrated in Figure \ref{fig:conv_top}. We refer to this as the Convergent Coupling (CC) model. 

We obtain the following equations: for $B_{1}$ and $B_2$
\begin{equation}\label{eq_st:tau1,2}
\tau_{d,i}
=
\rho_I g \frac{H_i}{L_i}
\left(
H_i
-
\frac{W_i}{W_3} H_3
\right),
\end{equation}
where the $B_1$ and $B_2$ ice thickness state equations follow equation (\ref{eq:H}). For $B_3$
\begin{equation}\label{eq_st:H3}
    \frac{dH_3}{dt} 
    =
     a_c - \frac{H_3 u_{b,3}}{L_3} + \frac{W_1}{W_3}\frac{H_1 u_{b,1}}{L_3} + \frac{W_2}{W_3}\frac{H_2 u_{b,2}}{L_3} ,
\end{equation}
where the $B_3$ driving stress equation follows (\ref{eq:tau_d}). Note that in this configuration $W_1 + W_2 = W_3$.

\begin{figure}[!htbp]
    \centering
    \includegraphics[width=\textwidth]{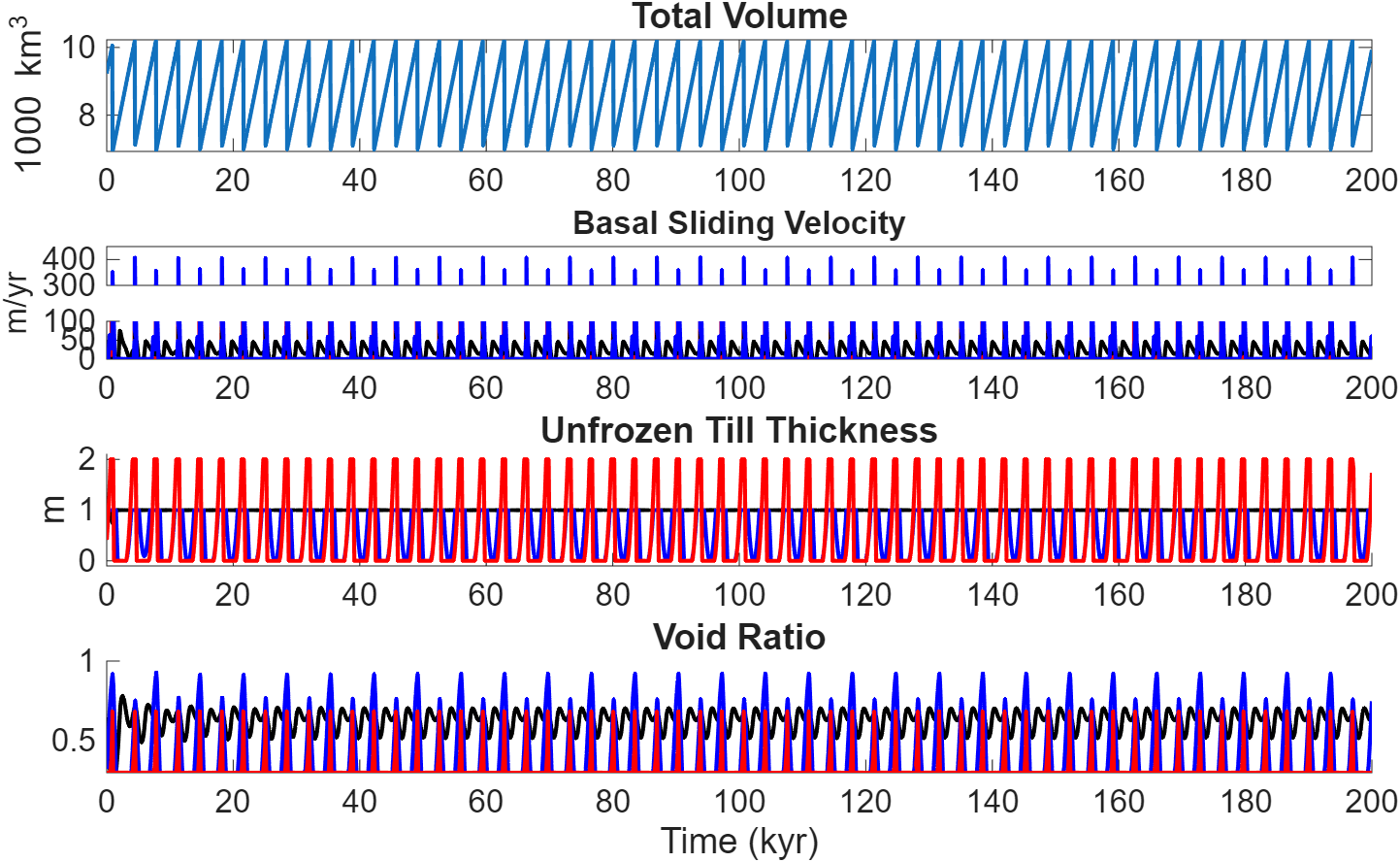}
    \caption{Diagnostic time series of four variables demonstrating periodic variability in the CC model. $L_2 = 190.8$km, $T_{s,2} = - 15^\circ \text{C}$, other parameters as in Table \ref{table:constants}. $B_1$ in black, $B_2$ in blue, $B_3$ in red.}
    \label{fig:shared_terminus_ts}
\end{figure}

\begin{figure}[!htbp]
    \centering
    \begin{subfigure}[b]{0.5\textwidth}
        \centering
        \includegraphics[width=\textwidth]{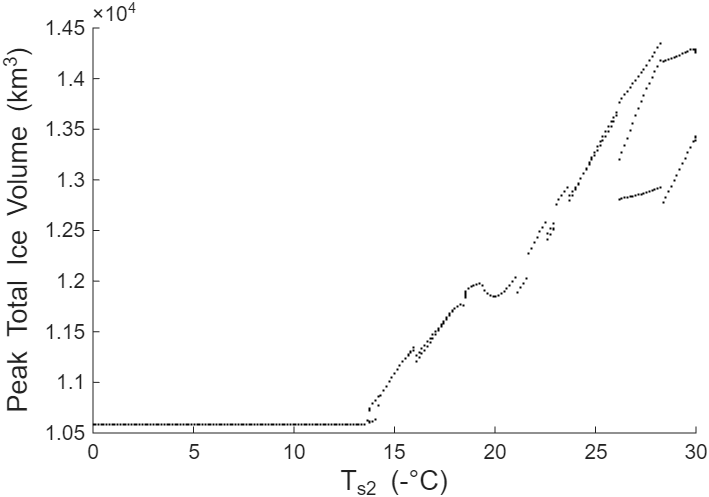}
        \caption{}
        \label{fig:shared_terminus_temp}
    \end{subfigure}
    \hfill
    \begin{subfigure}[b]{0.44\textwidth}
        \centering
        \includegraphics[width=\textwidth]{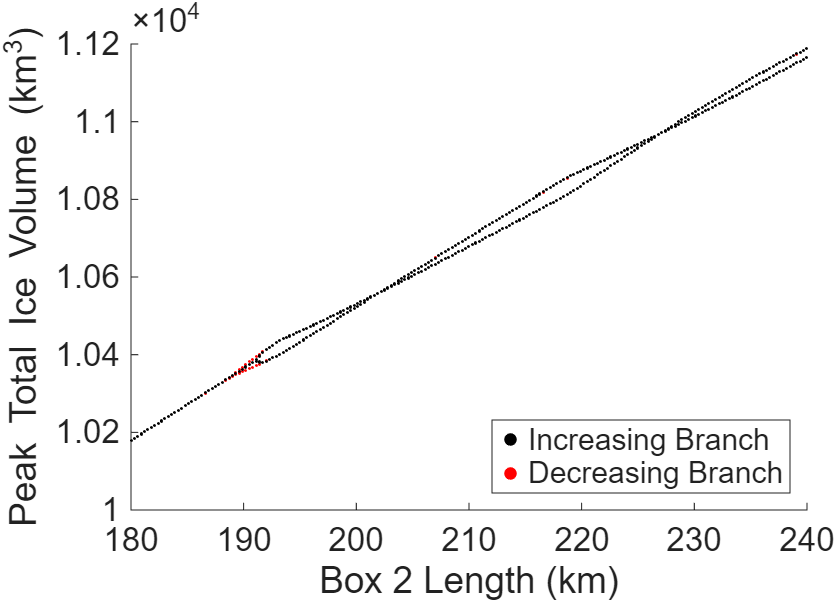}
        \caption{}
        \label{fig:shared_terminus_L2}
    \end{subfigure}

    \caption{Bifurcation diagrams for the CC model, neither demonstrate non-periodic behaviour. (a) $T_{s,2}$ is varied from $0^\circ$C to $-30^\circ$C; (b) $L_2$ increased and decreased from $180$km to $240$km. Other parameters as in Table \ref{table:constants}}
    \label{fig:shared_terminus_bifs}
\end{figure}

To explore the potential for chaotic behaviour in the CC model, both time series and bifurcation diagrams varying $T_{s,2}$ and $L_2$ were created using the same methods as for the DC model, with parameters adjusted to account for the change in topology (see Table \ref{table:constants}). Figure \ref{fig:shared_terminus_ts} demonstrates typical behaviour for this model, such that periodic oscillations in all three boxes are synchronous. The bifurcation diagrams in Figure \ref{fig:shared_terminus_bifs} also suggest this, as there is neither the presence of non-periodic behaviour, nor notable hysteresis.  

 While an absence of chaotic behaviour does not imply it cannot exist in this configuration, it suggests that there is no mechanism for chaos due to the different relationship between the ice streams.

In the DC model (see Figure \ref{fig:div_top}), chaotic behaviour occurs as a result of two or more frequencies becoming incommensurate. This is the result of the downstream boxes $B_2$ and $B_3$ both drawing ice from the shared reservoir which is $B_1$, creating a cross-coupled system where varying the parametrisation of $B_2$ changes the driving stress and thus the basal velocity of $B_1$, modifying the incoming mass of $B_3$.

It is hard to see how this can occur in a convergent topology (see Figure \ref{fig:conv_top}), because, if $B_1$ becomes incommensurate, it cannot modify the amount of volume entering $B_2$, as the boxes only share a terminus. An incommensurate $B_1$ would only result in modifying the driving stress of $B_2$ through the change in volume added to the shared terminus $B_3$. However, this change in driving stress is uniform across both $B_1$ and $B_2$, and as such would not result in a feedback loop creating chaotic behaviour. 

In contrast, Sayag \textit{et al.} \citep{sayag2011} present a spatially resolved 2D continuum model that demonstrates how seemingly chaotic variability can emerge in topologies similar to both the DC and CC models, due to local instabilities as a result of different subglacial topologies. While the approach is different, many qualitative behaviours, such as multiple flow regimes, bi-stability and chaos via resource competition from an upstream reservoir, are shared between the two frameworks. However, Sayag \textit{et al.} rely on local spatial interactions and lateral shear to generate instability, whereas the spatially lumped DC model demonstrates that a divergent topology alone provides a sufficient macroscopic mechanism for chaos, due to the shared reservoir. 

\section{Conclusions}
\label{sec:discuss}

In summary, we correct the K26 model of coupled ice streams in a divergent topology to ensure volume conservation, investigate the temporal dynamics of a corrected Divergent Coupling (DC) model and compare this with an equivalent Convergent Coupling (CC) model. Both models demonstrate the same variability as in R13, allowing for both binge-purge and steady streaming behaviour. The CC model exhibits only periodic behaviour, as in R13, while the DC model can exhibit both chaos and synchronisation. Strange attractors emerge through torus breakdown, appearing in intermittent chaotic windows between the locked states of a period-adding cascade. We observe hysteresis and chaotic transients when increasing ice stream length similar to K26, as well as when decreasing ice stream length, further supporting the findings of \cite{K26b}. The DC model exhibits Arnold tongues, suggesting synchronisation between ice streams may greatly impact variability regardless of timescale. 

Chaotic behaviour was found to be rare across the parameter regimes tested, with the most chaotic behaviour demonstrating a Largest Lyapunov Exponent of $\lambda \approx 0.3$, which corresponds to an error-doubling time of $\sim 7\,\text{kyr}$. As this occurs on the same timescale as Heinrich Events it supports the hypothesis that chaotic variability may contribute to the irregular alignment of Dansgaard-Oeschger events with Heinrich events \citep{macayeal_bingepurge_1993,K26}. 
Synchronisation between divergent ice streams has been shown to create phase locking and Arnold tongues. Despite chaotic behaviour limiting predictability on the scale of thousands of years, the presence of chaotic transients and synchronisation implies ice streams can be highly variable regardless of timescale. 

The R13 model, and consequently the DC and CC models, are strongly idealised in geometry and the processes modelled. However, ice streams are fundamentally connected to a wider geophysical system, and other work has shown that a divergent topology is not a requirement for synchronisation. For example, the Arnold tongues found in \citep{MannRobel} occur due to interaction of an ice stream with submarine melt at the grounding line and freshwater entering the ocean. Similarly, marine ice sheet instability (MISI) modified by glacial isostatic adjustment can result in ice sheet variability on paleoclimate timescales \cite{feldmann_rate-induced_2025}. Properly understanding these and other ice-ocean interactions is important for evaluating how anthropogenic climate change drives ice sheet changes and variability, and the resulting sea level rise \citep{reese_far_2018}.


Several other extensions to R13 have been proposed, such as noisy atmospheric forcing \citep{mantelli_stochastic_2016}, alongside the melt-elevation feedback and finite propagation timescale for the vertical temperature gradient highlighted in \citep{K26}. 
Such additions are essential for understanding ice sheet tipping and Heinrich event variability, and we hypothesise that noise-induced tipping could arise from anthropogenic climate change increasing the amplitude of atmospheric variability.

Of more direct relevance to the mechanism identified here is the sub-temperate sliding formulation of \cite{mann_subtemperate_2025}, which removes the piecewise continuous nature of basal sliding in R13. However, the dynamical switches which we show admit chaotic behaviour remain, thus we expect both chaos and Arnold tongues to persist in a formulation of divergent ice streams with sub-temperate sliding. Testing this, and exploring the interaction of these extended formulations with the DC model more broadly poses interesting future work.

The Siple Coast Ice Streams (SCIS) offer geographical analogues for both topologies: Mercer and Whillans share a terminus, as in the convergent case, while Kamb and Bindschadler share an upstream reservoir, as in the divergent case. Kamb and Bindschadler stagnated $\sim 170$ and $\sim450$ years ago, respectively, with Bindschadler retreating and reactivating before Kamb stagnated \citep{conway2002}, whereas Mercer and Whillans last retreated over $\sim 4000$ years ago \citep{Neuhaus}. This is consistent with the behaviour demonstrated in this work, suggesting coupling topology has a significant effect on the mechanism behind these contrasting dynamics despite otherwise similar environmental conditions. As such, further work could consist of creating a model of the full SCIS system to further analyse this mechanism.

Box models fix many aspects (such as the coupling topology and base areas) by construction, so establishing whether the coupling effects identified here persist under realistic geometries requires ice sheet models that resolve these, such as Yelmo \citep{robinson_description_2020}. Modelling the SCIS with such a model would be a key test for the impact of convergent and divergent topologies. Additionally, modelling a theoretical Hudson Strait Ice Stream would allow one to test the hypothesis that the irregular timing of Dansgaard-Oeschger events with Heinrich events is influenced by chaotic ice stream variability \citep{R13}.

\begin{acknowledgements}

The authors would like to thank Alexander Robel and Kolja Kypke for helpful conversations regarding their models, and the authors of Kypke \textit{et al.} \cite{K26} for allowing their figures and code to be used. JG also thanks Ian Hewitt for the opportunity to present this work to his `Icy Maths' research group and receive feedback regarding the volume conservation error this work addresses. Moreover,
JG would like to thank the Professor Ozgur Akman Memorial Bursary for funding the opportunity to speak at BAMC 2026 and the aforementioned presentation to Ian Hewitt’s research group. The research of TS was partly supported by the European Union’s Horizon Europe research and innovation programme under the Marie Sklodowska-Curie grant agreement ErgodicHyperbolic - 10115118. The research of PA was partly supported by the Horizon Europe grant agreement Past2Future (Grant No. 101184070).
For the purpose of open access, the author has applied a Creative Commons Attribution (CC BY) licence to any Author Accepted Manuscript version arising from this submission.
\end{acknowledgements}

\section*{Code availability}

The code used to conduct the analysis and create the figures in this body of work is available: \url{https://github.com/JoshuaGrimstead/Coupled-Ice-Streams/}

\appendix

\section{Parameter tables} \label{appendix:param}
Empirical constants used in this work for the DC and CC models are shown in Table \ref{table:empirical}, with state variables recorded in Table \ref{table:state}. Parametrisations for other variables are shown in Table \ref{table:constants}.

\begin{table*}[!htbp] \caption{Constants with empirical values used in the Divergent Coupling and Convergent Coupling models.}
\centering
\renewcommand{\arraystretch}{1.2}
\begin{ruledtabular}
\begin{tabular}{llccl}
Symbol & Description & Prescribed value & Units \\
\hline
$\tau_0$ & Empirical till coefficient & $9.44\times10^8$ & Pa \\
$c$ & Empirical till exponent & 21.7 & — \\
$A_g$ & Glen's law rate factor & $5\times10^{-25}$ & Pa$^{-3}$ s$^{-1}$ \\
$n$ & Glen's law exponent & 3 & — \\
$q_g$ & Geothermal heat flux & 0.07 & W m$^{-2}$ \\
$g$ & Acceleration due to gravity & 9.81 & m s$^{-2}$ \\
$\rho_I$ & Density of ice & 917 & kg m$^{-3}$ \\
$K_I$ & Thermal conductivity of ice & 2.1 & J s$^{-1}$ m$^{-1}$ K$^{-1}$ \\
$C_I$ & Volumetric heat capacity of ice & $1.94\times10^6$ & J K$^{-1}$ m$^{-3}$ \\
$L_f$ & Latent heat of fusion & $3.335\times10^5$ & J kg$^{-1}$ \\
\end{tabular}
\end{ruledtabular}
\label{table:empirical}
\end{table*}

\begin{table*}[!htbp] \caption{State variables evolved by the Divergent Coupling and Convergent Coupling models}
\centering
\renewcommand{\arraystretch}{1.2}
\begin{ruledtabular}
\begin{tabular}{llccl}

Symbol & Description & Units \\

\hline
$H_i$ & Ice thickness in box $i$ & m \\
$T_{b,i}$ & Basal temperature in box $i$  & $-^\circ$C  \\
$e_i$ & Till void ratio in box $i$  & —  \\
$u_{b,i}$ & Basal sliding velocity in box $i$   & m s$^{-1}$  \\
$\tau_{d,i}$ & Driving stress in box $i$ & Pa  \\
$\tau_{f,i}$ & Basal frictional stress in box $i$ & Pa \\
$h_{\mathrm{till},s,i}$ & Thickness of solids in the unfrozen till & m \\

\end{tabular}
\end{ruledtabular}
\label{table:state}
\end{table*}

\begin{table*}[!htbp]
\caption{\label{table:constants}%
Prescribed constants used in the Divergent Coupling and Convergent Coupling models.
Rows grouped under one symbol are box-specific ($B_1$, $B_2$, $B_3$). Constants that are changed from base value are indicated using $^*$ with the alternative values being indicated in the caption of the respective plots.}
\begin{ruledtabular}
\begin{tabular}{llccl}
Symbol & Description & Converging & Diverging & Units \\
\hline
$e_c$ & Till consolidation threshold & \multicolumn{2}{c}{$0.3$} & \textemdash \\
$\eta_b$ & Thickness of temperate ice layer & \multicolumn{2}{c}{$10$} & m \\
$a_{c,i}$ & Accumulation rate & \multicolumn{2}{c}{$0.05$} & m\,yr$^{-1}$ \\
 & & \multicolumn{2}{c}{$0.0455$} & \\
 & & \multicolumn{2}{c}{$0.0417$} & \\
$h_{\mathrm{till,max},i}$ & Maximum unfrozen till solid thickness & $1$ & $1$ & m \\
 & & $1$ & $2$ & \\
 & & $2$ & $2$ & \\
$T_{s,i}$ & Surface temperature in box $i$ & \multicolumn{2}{c}{$15$} & $^\circ$C \\
 & & \multicolumn{2}{c}{$14.02^{*}$} & \\
 & & \multicolumn{2}{c}{$15$} & \\
$L_i$ & Ice stream trunk length in box $i$ & $200$ & $55^{*}$ & km \\
 & & $215^{*}$ & $215^{*}$ & \\
 & & $135$ & $250$ & \\
$W_i$ & Ice stream trunk width in box $i$ & $25$ & $60$ & km \\
 & & $35$ & $35$ & \\
 & & $60$ & $25$ & \\
\end{tabular}
\end{ruledtabular}
\end{table*}

\section{Validating volume conservation}

\subsection{The error in the Kypke et al. formulation} \label{appendix:K26_error}
The formulation of a three box topology relies on ice volume being conserved between the upstream and downstream boxes, such that no volume is gained from or lost to other sources. We demonstrate how volume is not conserved between the upstream and downstream boxes in the K26 divergent formulation by converting the ice stream thickness equations into volume equations. 

Starting with equation (\ref{eq:K26_H2,3}) in the case where $\frac{dV_1}{dt} < 0$, which describes volume flux in $B_2$ and $B_3$
\begin{equation*}
L_i W_i \frac{dH_i}{dt} = L_i W_i a_c - W_i H_i u_{b,i} - \frac{W_i}{W_1} \frac{dV_1}{dt}, \quad i=2,3,
\end{equation*}

 where volume flux from $B_1$ is $\frac{dV_1}{dt} = L_1 W_1 a_c - (W_1 H_1 - W_2 H_2 - W_3 H_3) u_{b,1}$. To simplify, we define the internal flux leaving $B_1$ as 
\begin{equation*}
Q_f = (W_1 H_1 - W_2 H_2 - W_3 H_3) u_{b,1},
\end{equation*}
such that the volume change in $B_1$ is $\frac{dV_1}{dt} = L_1 W_1 a_c - Q_f$. Substituting this back into equation (\ref{eq:K26_H2,3}) gives:
\begin{equation*}
L_i W_i \frac{dH_i}{dt} = L_i W_i a_c - W_i H_i u_{b,i} - \frac{W_i}{W_1} (L_1 W_1 a_c - Q_f).
\end{equation*}
Adding the equations for $i=2,3$ (noting that $L_i W_i \frac{dH_i}{dt} = \frac{dV_i}{dt}$) we obtain:
\begin{align*}
\frac{dV_2}{dt} + \frac{dV_3}{dt} &= a_c(L_2W_2 + L_3W_3) - W_2H_2u_{b,2} - W_3H_3u_{b,3} \\
&\quad - \frac{W_2+W_3}{W_1} (L_1 W_1 a_c - Q_f).
\end{align*}
Using $W_1 = W_2 + W_3$, the fractional multiplier becomes 1, leaving
\begin{align*}
\frac{dV_2}{dt} + \frac{dV_3}{dt} &= a_c(L_2W_2 + L_3W_3) - W_2H_2u_{b,2} - W_3H_3u_{b,3} \\
&\quad - L_1W_1a_c + Q_f.
\end{align*}
Adding the volume change of $B_1$ ($\frac{dV_1}{dt} = L_1 W_1 a_c - Q_f$) we find the total global volume change
\begin{align*}
\frac{d}{dt}(V_1+V_2+V_3) &= a_c(L_2W_2 + L_3W_3) - W_2H_2u_{b,2} - W_3H_3u_{b,3} \\
&\quad - L_1W_1a_c + Q_f + L_1W_1a_c - Q_f,
\end{align*}
which simplifies to
\begin{equation*}
\frac{d}{dt}(V_1+V_2+V_3) = a_c(L_2W_2+L_3W_3) - W_2H_2u_{b,2} - W_3H_3u_{b,3}.
\end{equation*}
This is crucially missing the accumulation on the area of $B_1$, $a_c L_1 W_1$. 

In addition, the switching in equation (\ref{eq:K26_H2,3}) is unrealistic. When $\frac{dV_1}{dt} > 0$ accumulation $a_c$ will dominate such that $B_1$ continues to lose volume. This is omitted in equation (\ref{eq:K26_H2,3}), thus volume conservation is not supported in either case.

\subsection{Corrected divergent topology}\label{appendix:div_fix}
From ice thickness equations (\ref{eq:H1_fix}) and (\ref{eq:H2,3_fix}) substituted into
\[
\frac{dV_i}{dt}=L_iW_i\frac{dH_i}{dt}, \qquad i=1,2,3,
\]
the volume equations are

\begin{align}
\frac{dV_1}{dt} &= W_1L_1a_c-W_1H_1u_{b,1}, \label{eq:dV1}\\
\frac{dV_2}{dt} &= W_2L_2a_c-W_2H_2u_{b,2}+W_2H_1u_{b,1}, \label{eq:dV2}\\
\frac{dV_3}{dt} &= W_3L_3a_c-W_3H_3u_{b,3}+W_3H_1u_{b,1}. \label{eq:dV3}
\end{align}
Since $W_1=W_2+W_3$, the outflow from $B_1$ equals the combined inflow to $B_2$ and $B_3$. Summing the box volumes
\begin{equation*}
\frac{dV_{\mathrm{tot}}}{dt}
=\frac{dV_1}{dt}+\frac{dV_2}{dt}+\frac{dV_3}{dt},
\label{eq:vol_cons}
\end{equation*}
and substituting (\ref{eq:dV1})--(\ref{eq:dV3}) gives

\begin{align}
\frac{dV_{\mathrm{tot}}}{dt}
&=W_1L_1a_c+(-W_1+W_2+W_3)H_1u_{b,1}\nonumber\\
&\quad+W_2L_2a_c-W_2H_2u_{b,2}
+W_3L_3a_c-W_3H_3u_{b,3}.
\end{align}
Using $W_1=W_2+W_3$, the internal fluxes cancel leaving

\[
\frac{dV_{\mathrm{tot}}}{dt}
=W_1L_1a_c+W_2L_2a_c+W_3L_3a_c
-W_2H_2u_{b,2}-W_3H_3u_{b,3},
\]
which is total accumulation minus the outgoing terminus flux.

\vspace{1em}

\subsection{Convergent topology} \label{appendix:conv}

Similar to the divergent topology,
\[
\frac{dV_i}{dt}=L_iW_i\frac{dH_i}{dt},
\]
thus
\begin{equation*}
\frac{dV_{\mathrm{tot}}}{dt}
=\frac{dV_1}{dt}+\frac{dV_2}{dt}+\frac{dV_3}{dt}.
\label{eq:vol_cons_st}
\end{equation*}
Substituting (\ref{eq:H}) and (\ref{eq_st:H3}) yields
\begin{align}
\frac{dV_{\mathrm{tot}}}{dt}
&=L_1W_1\left(a_c-\frac{H_1u_{b,1}}{L_1}\right)\nonumber\\
&\quad+L_2W_2\left(a_c-\frac{H_2u_{b,2}}{L_2}\right)\nonumber\\
&\quad+L_3W_3\left(a_c-\frac{H_3u_{b,3}}{L_3}
+\frac{W_1}{W_3}\frac{H_1u_{b,1}}{L_3}
+\frac{W_2}{W_3}\frac{H_2u_{b,2}}{L_3}\right),
\end{align}
which simplifies to
\begin{align}
\frac{dV_{\mathrm{tot}}}{dt}
&=\left(a_cL_1W_1-W_1H_1u_{b,1}\right)\nonumber\\
&\quad+\left(a_cL_2W_2-W_2H_2u_{b,2}\right)\nonumber\\
&\quad+\left(a_cL_3W_3-W_3H_3u_{b,3}
+W_1H_1u_{b,1}+W_2H_2u_{b,2}\right).
\end{align}
The internal fluxes cancel, giving
\begin{equation*}
\frac{dV_{\mathrm{tot}}}{dt}
=a_c(L_1W_1+L_2W_2+L_3W_3)-W_3H_3u_{b,3},
\label{eq:vol_cons_tot_final}
\end{equation*}
which is total accumulation minus the downstream outflow, demonstrating volume conservation.

\section{Applying Rosenstein's algorithm for estimating Largest Lyapunov Exponents to Poincaré sections}\label{appendix:Rosn}

\subsection{Model limitations and preliminary adjustments}

For an $n$-dimensional system, divergence or convergence can occur in any of the $n$ distinct directions of the phase space, resulting in $n$ different LEs creating a Lyapunov spectrum $\lambda_1 \ge \lambda_2 \ge \dots \ge \lambda_n$. However, as there are discontinuities in the derivative of the model we can only estimate the Largest Lyapunov Exponent (LLE) from phase space which appears continuous.

We estimate the Largest Lyapunov Exponent (LLE) by creating 11D Poincaré sections from 12D phase space, saving the full system as $e_2$ increases through $0.6$ and comparing successive sections to track the divergence of nearby trajectories. 

However, not all state space dimensions benefit the estimation of chaotic behaviour. As $B_1$ is always in the steady streaming mode for these estimations, it has an arbitrarily large void ratio $e_1$ which distorts true attractor size. Consequently, the box 1 basal temperature $T_{b,1}$ and till thickness $h_{\mathrm{till},s,1}$ do not evolve (although $H_1$ can evolve due to the coupled driving stress equation buttressing $B_1$). Similarly, $B_3$ cannot leave the weak binge-purge mode such that $h_{\mathrm{till},s,3}$ and $T_{b,3}$ cannot evolve. Thus we only track the evolution of the 6D phase space composed of $H_1$, $H_2$, $h_{\mathrm{till},s,1}$, $T_{s,2}$, $H_3$ and $e_3$, which reduces the impact of solver noise on LLE estimation.

\subsection{Rosenstein's algorithm}

Rosenstein \textit{et al.} \cite{rosenstein_practical_1993} give a practical algorithm for extracting LLEs from noisy data. While the original algorithm uses time series and reconstructs a state space using Takens' theorem, as we are able to obtain Poincaré sections this is unnecessary. The algorithm described by \cite{rosenstein_practical_1993} takes a point in phase space, pairs this with its nearest neighbour in that space (excluding neighbours that are close
because they are only close in time) and estimates the separation of each pair across time steps. Writing $d_i(k)$ for the separation of the $i$th pair after $k$ steps, the algorithm computes
\begin{equation*}
L(k) = \big\langle \ln d_i(k) \big\rangle_i 
\end{equation*}
where $\langle\cdot\rangle_i$ denotes the arithmetic mean over all pairs $i$, such that $L(k)$ is the mean logarithmic separation across all pairs.
For a chaotic system trajectories separate, following $d_i(k) \sim d_i(0)e^{\lambda k}$, until exponential separation ceases such that $L(k)$ is linear with slope $\lambda$; the LLE estimated using least-squares fitting over that linear region. Note that no embedding dimension or delay needs to be chosen for this method as the Poincaré sections represent true dynamics, as opposed to a reconstruction.

\subsection{Ensuring robust LLE estimations}

After excluding unnecessary dimensions, the remaining dimensions are normalised by the mean before distances are computed, as each parameter has its own scale which can vary by several orders of magnitude. After normalisation, the diameter of the attractor is estimated using the 90th percentile of pairwise distances, denoted by $D$, rather than the full diameter to reduce the influence of isolated outliers. This characteristic scale is then used to define the fitting window for the LLE estimate. Because the attractor has a finite diameter, separations cannot grow beyond the true attractor diameter and $L(k)$ must eventually saturate regardless of the dynamics. The fit is therefore terminated when $L(k)$ first reaches $\ln{(\gamma D)}$, choosing $\gamma=0.5$. In addition, we use the attractor diameter to isolate period-1 regions from the remaining dynamics, as after normalisation numerical noise in a small attractor can lead to false positives when detecting chaos. As such we use this identifier of period-1 behaviour to mask the LLE plots.

In order to ensure that neighbouring points that are close in time are not used to track the divergence of two trajectories (being part of the same trajectory), we impose a Theiler window; the minimum number of iterates (or lag) separating neighbour pairs \citep{theiler_spurious_1986}. This window is estimated independently for each parameter cell as the first lag at which the autocorrelation of the Poincaré section sequence falls below $e^{-1}$, providing an estimate of the decorrelation time. We neglect the initial separation $d_i(0)$, as the nearest neighbour is arbitrarily close and produces an artificially large separation rate. Thus we initiate the least-squares fitting from $k=1$. 
\\

\bibliographystyle{abbrvnat}
\bibliography{MSc_Dissertation}

\end{document}